%% file: main.tex
\newif\ifusenix
\newif\ifacm
\newif\ifmcom

\usenixfalse
\acmtrue
\mcomfalse

\ifusenix
 \documentclass[letterpaper,twocolumn,10pt]{article}
\usepackage{usenix2019_v3}
\usepackage{times}
\fi
\ifacm
  \documentclass[10pt,sigconf]{acmart}
  \renewcommand\footnotetextcopyrightpermission[1]{} % removes footnote with conference info
  \setcopyright{none}
\fi

\ifmcom
  \documentclass{sig-alternate-10pt}
\fi

\usepackage{booktabs} %added by Ish for nice looking tables
\usepackage{pifont}% http://ctan.org/pkg/pifont for tick and cross symbols
\usepackage{xcolor}
\usepackage{amsfonts}
\usepackage{balance}
\PassOptionsToPackage{hyphens}{url}
\usepackage{hyperref}
\usepackage{color}
\usepackage{graphics}
\usepackage{graphicx}
\usepackage{url}
\usepackage{listings}
\usepackage{multicol}
\usepackage{multirow}
\usepackage[scaled]{helvet}
\usepackage{rotating}
\usepackage{xspace}
\usepackage{algorithm}
\usepackage{comment}
\usepackage{enumitem}
\usepackage{amsmath}
\usepackage{mathrsfs}
\usepackage{cancel}
\usepackage{cleveref}
\usepackage{subcaption}
\usepackage{commath}
\usepackage{enumitem}

\usepackage[labelfont=bf]{caption}
\usepackage[explicit,noindentafter]{titlesec}

\titlespacing\section{4pt}{2pt plus 2pt minus 2pt}{3pt plus 2pt minus 2pt}
\titlespacing\subsection{4pt}{3pt plus 2pt minus 2pt}{3pt plus 2pt minus 2pt}
\titlespacing\subsubsection{4pt}{3pt plus 2pt minus 2pt}{3pt plus 2pt minus 2pt}

\newcommand{\newtodo}[1]{\ClassWarning{NOT READY TO SUBMIT}{There is something left todo} \textcolor{blue}{}}

\newcommand{\review}[1]{\ClassWarning{NOT READY TO SUBMIT}{There is something left todo} \textcolor{orange}{Reviewer 1}}%[Review: #1 ]

\newcommand{\name}{CommRad\xspace}

\copyrightyear{2024}
\acmYear{2024}
\setcopyright{acmcopyright}\acmConference[Arxiv '24]{Uploaded to Arxiv}{July 3, 2024}{USA}


\input{mmreliable_defines}

\begin{document}
% %%%%%%%%%% Agrim commands-2 start%%%%%%%%%%%%
% \interfootnotelinepenalty=10000
% \setlength{\belowdisplayskip}{2pt} \setlength{\belowdisplayshortskip}{2pt}
% \setlength{\abovedisplayskip}{2pt} \setlength{\abovedisplayshortskip}{2pt}
% %%%%%%%%%%%% end %%%%%%%%%%%%%%%%%%%%
% \title{Supporting multi-user with multi-beam multi-frequency delay-based phased array design}
\title{CommRad: Context-Aware Sensing-Driven Millimeter-Wave Networks
% CommRad: Context-Aware Sensing-Driven mmWave Networks
% Looking through two eyes: Scalable Sensing-Driven Collaborative mmWave System
% Scalable Sensing-Driven Mobile mmWave Communication System
%\name: Scalable and deployable multi-beam mmWave systems with environment sensing
% Scaling and Augmenting multi-beam 5G mmWave systems with environment sensing
}

% \author{Paper \#617 12 pages + References}

\author{Ish Kumar Jain, Suriyaa MM, Dinesh Bharadia}
\affiliation{
    \institution{University of California San Diego}
    % \city{La Jolla}
    % \state{CA}
    % \country{USA}
    }
\renewcommand{\shortauthors}{IK Jain, Suriyaa MM, D.Bharadia}
\email{{ikjain, smm233, dineshb}@ucsd.edu}

% \renewcommand{\shorttitle}{Collaborative Learning for Sensing-Driven mmWave Network}


\ifacm
    \input{0_abstract}
\fi

\maketitle

\ifusenix
    \input{0_abstract}
\fi

\ifmcom
    \input{0_abstract}
\fi

\input{1_intro_v7}
\input{2_related_v2}

\input{3_design_v4}
\input{4_implementation}

\input{6_evaluation_v3.tex}
\input{8_discussion}

% Bibliography
\newpage
\bibliographystyle{unsrt}
\balance
\bibliography{acmart, mmreliable}
\label{lastpage}
\end{document}

%% file: mmreliable_defines.tex
\definecolor{deepblue}{rgb}{0,0,0.5}
\definecolor{deepred}{rgb}{0.6,0,0}
\definecolor{deepgreen}{rgb}{0,0.5,0}
\definecolor{backcolour}{rgb}{0.95,0.95,0.92}

\def\beq{\begin{equation}}
\def\eeq{\end{equation}}
\def\beqa{\begin{eqnarray}}
\def\eeqa{\end{eqnarray}}
\def\beqan{\begin{eqnarray*}}
\def\eeqan{\end{eqnarray*}}

\def\x{\times}

\def\qed{\mbox{} \hfill $\Box$}
\def\tm1{t\! - \! 1}
\def\tp1{t\! + \! 1}

%% file: 0_abstract.tex
% !TEX root = main.tex

\begin{abstract}
Millimeter-wave (mmWave) technology is pivotal for next-generation wireless networks, enabling high-data-rate and low-latency applications such as autonomous vehicles and XR streaming. However, maintaining directional mmWave links in dynamic mobile environments is challenging due to mobility-induced disruptions and blockage. While effective, the current 5G NR beam training methods incur significant overhead and scalability issues in multi-user scenarios. To address this, we introduce \name, a sensing-driven solution incorporating a radar sensor at the base station to track mobile users and maintain directional beams even under blockages. While radar provides high-resolution object tracking, it suffers from a fundamental challenge of lack of context, i.e., it cannot discern which objects in the environment represent active users, reflectors, or blockers. To obtain this contextual awareness, \name unites wireless sensing capabilities of bi-static radio communication with the mono-static radar sensor, allowing radios to provide initial context to radar sensors. Subsequently, the radar aids in user tracking and sustains mobile links even in obstructed scenarios, resulting in robust and high-throughput directional connections for all mobile users at all times. We evaluate this collaborative radar-radio framework using a 28 GHz mmWave testbed integrated with a radar sensor in various indoor and outdoor scenarios, demonstrating a 2.5x improvement in median throughput compared to a non-collaborative baseline. 

\end{abstract}

%% file: 1_intro_v7.tex
% !TEX root = main.tex
\section{Introduction}\label{sec:intro}

% New in v3: Rewritten from scratch after missing Mobicom. After a talk at Qualcomm, some ideas are coming for Integrated sensing and communication. Group meeting helped too.
% New in v4: After Dinesh's feedback and before gpt- remove all previous comments

%new in v6: After Sigcomm'24 rejection, and group meeting on 6/28/24, rewriting the intro. Issue: Two motivation, one for mmwave and second for Sensing-driven comm. Discussed two options to resolve this - 1. Start with only mmwave motivation and requirements without telling sensing driven comm and then bring sensing later. Option 2. Start with selling sensing and telling that it is already a norm and great, everyone using it, but it has some problem that we fix it. The issue with option 1 is it takes too late to introduce what we doing. Issue with opt 2 is it does not tell what problem in mmwave are we using. Finally we decide for option 1.5 - take first half of option 1: mmweave requirements. Them move directly to sensing-driven comm, without diverting to why radio alone is not enough. 

%new in v7: incorporating Dinesh's feedback in first 3 paragraphs

% mmwave is great, mobile access is important, last mile connectivity. 

%para2 why radar is a great choice, how people are already using radars for many applications. 

% Purely base station change and no change for the mobile devices, no information needed from devices e.g. gps, IMU data. 

Millimeter-wave access is vital for the next-generation wireless network to connect the last mile devices with extremely high multi-Gbps data rates with ultra-low latency \cite{3gpp, fr2bands}. With such high data rates, future autonomous vehicles, loaded with sensors such as Lidar, radars, or cameras, can share their data with the infrastructure or other vehicles for safer driving; the wireless AR/VR devices would enjoy high-quality XR streaming for an immersive experience~\cite{abari2017enabling}. However, a major challenge in supporting these use cases is user mobility -- anything mobile is an issue for mmWave links. As millimeter-wave links deploy hundreds of antennas to create narrow beams to increase coverage range, they inadvertently fail to maintain these directional links in mobile scenarios. When the target users are mobile, the link requires tracking the users to maintain directional connectivity. Or when other objects in the environment are mobile, they can obstruct or block communication when they enter across an established link. 
While there are procedures for beam maintenance in 5G NR, such as beam training, where radio scans multiple beams to find the best direct or reflected-path beam for the users, these techniques suffer from high overhead manifested in their long duration and frequent repetitions. The more time a radio spends in beam training, the less time is available for communication, resulting in a high overhead of up to 25\% (5 ms duration of beam training repeated every 20 ms, according to 5G NR protocol~\cite{5gnr}). To make it worse, the overhead increases with the number of users, as each user demands separate time and frequency resources for feedback in the form of CSI reports, thus lacking scalability in multi-user scenarios.

\begin{figure}[t]
    \centering
\includegraphics[width=0.48\textwidth]{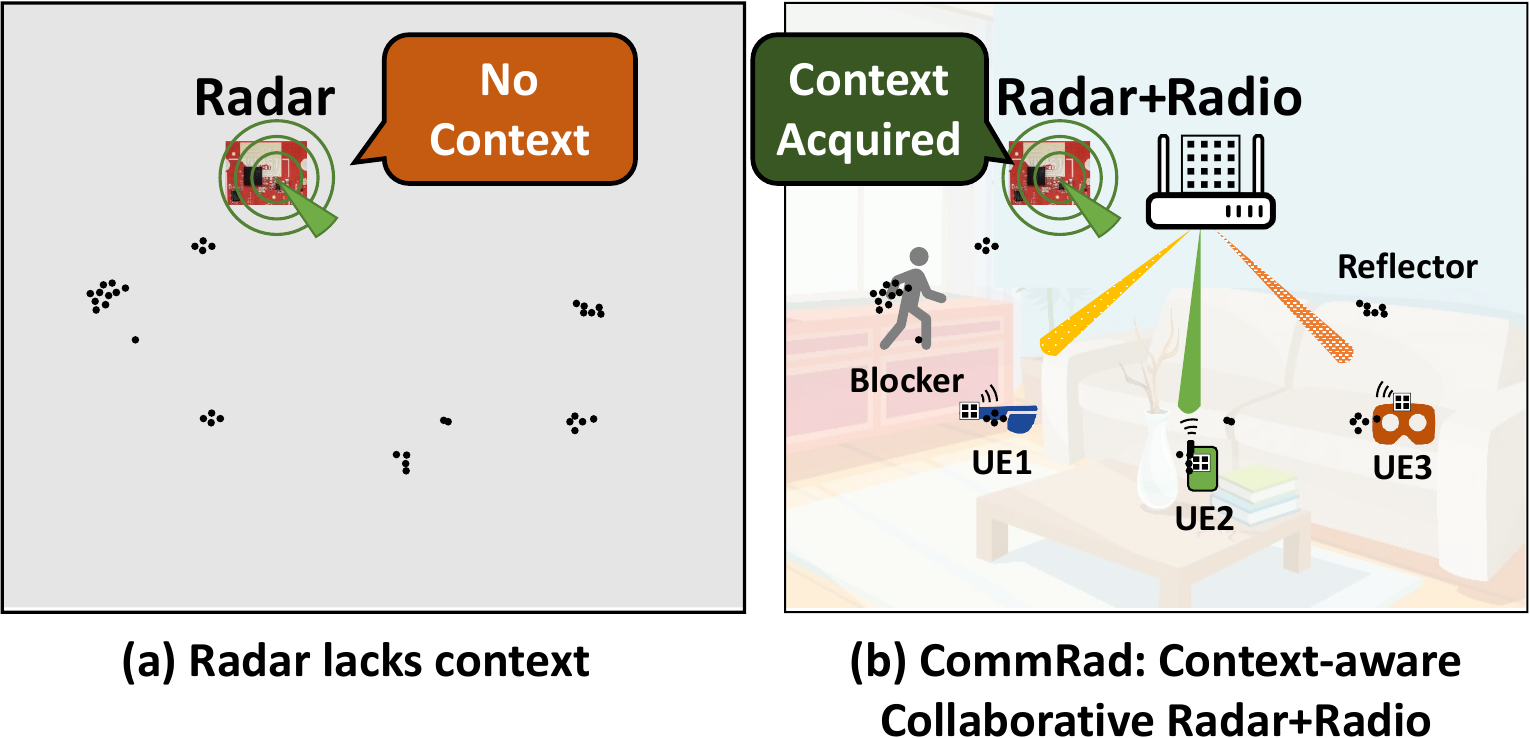}
    \caption{{\name is a collaborative learning framework for context-aware sensing-driven mmWave communication.}}
    \label{fig:prototype}
    % \vspace{-0.2in}
\end{figure}

To overcome the overhead, sensing-driven communication is envisioned as a pivotal technology where a dedicated sensor is deployed to aid in beam maintenance. The sensors can track users, identify good reflectors, proactively avoid environmental blockers, and instruct the radio about the best direct or reflected-path beam at a given time. Among many choices of sensors, a millimeter-wave radar sensor stands out for low-cost, privacy-preserving, and weather-independent sensing modality in both indoor and outdoor scenarios~\cite{zheng2023neuroradar, cui2022integrated, bansal2020pointillism, zheng2024enhancing, dunna2023r}. However, radar-sensor-driven communication suffers from a fundamental cold start problem as depicted in Figure~\ref{fig:prototype}. The problem is that the radar yields extensive reflections from many objects in the environment without having any context for those objects -- which objects are mmWave users that need to be tracked, or which objects are reflectors/blockers that can assist or obstruct communication links? Due to this contextual deficiency, prior implementation of radar-aided communication is limited to a straightforward clutter-free environment with only one user and does not support reflector-assisted links~\cite{demirhan2022radar,luo2023millimeter,aydogdu2020distributed, chen2023beam,chen2023radar}. These works rely on data-driven machine learning techniques to perform end-to-end training from the radar data to predict communication beam patterns directly, skipping the important step of acquiring context. The hope is that the model will somehow learn the context by feeding extensive data while training. 
% These models demand re-training for diverse environments, often involving labor-intensive data collection efforts with ground truth information, yet may still fall short of obtaining the requisite contextual awareness. 
To the best of our knowledge, no prior work on radar-driven communication experimentally demonstrated context-aware multi-user tracking while supporting a reflected-path link dynamically and blockage adaptation.

\begin{figure}
    \centering
    \includegraphics[width=.45\textwidth]{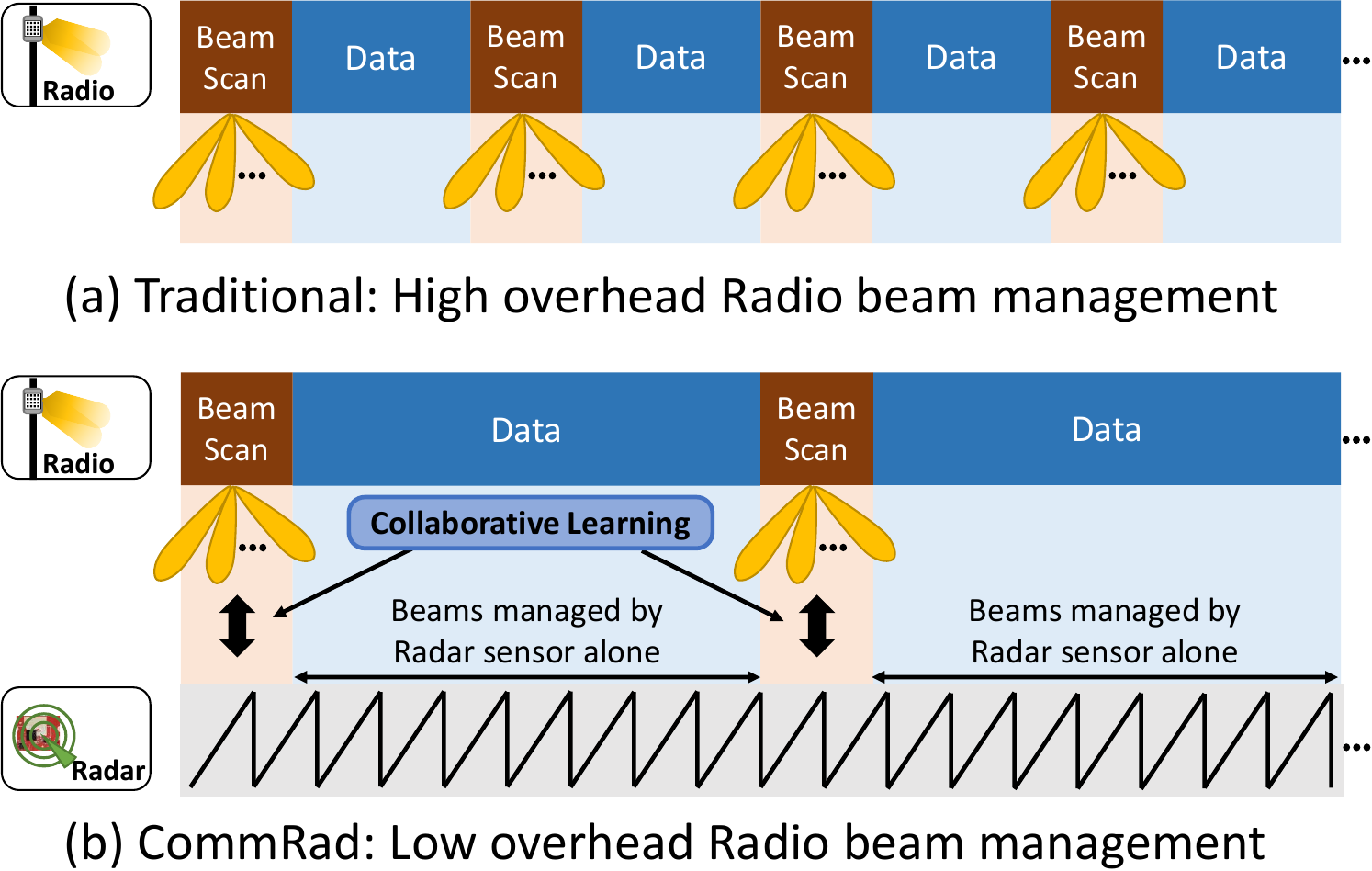}
    \caption{\name's bi-directional radar-radio collaborative learning framework to improve data communication efficiency by reducing radio beam scan overhead.}
    \label{fig:radio_radar_waveform}
\end{figure}

\textbf{\name:} This paper presents the design and implementation of \name, a sensing-driven communication framework driven by a fundamental insight: the communication radio can serve as a complementary sensing modality, providing the contextual awareness that radar sensing currently lacks. \name represents a \textbf{\textit{collaborative bi-directional learning framework}} that seamlessly integrates wireless sensing (radar) and the communication radio as sensing modalities to collectively comprehend the environment and user mobility within a relevant context. In this framework, we first deploy the radio as a sensor to provide the required context to the radar through the mandatory beam training procedure. We then deploy the context-aware radar sensor to track any movement of objects and aid in maintaining the beams while the radio is in the communication phase, as shown in Figure~\ref{fig:radio_radar_waveform}. This bi-directional radar-radio collaborative learning reduces the overhead due to frequent beam training and ensures that radar has appropriate context to maintain the beams, ensuring reliable and high throughput mobile mmWave links.

A key observation within \name is the recognition that the radio and radar units provide \textit{`complementary information'} essential for comprehending the environment and user mobility.
    \textbf{Radar as a Sensor:} The radar unit measures reflections from the environment and users through \emph{mono-static sensing}, where the radar Rx is co-located and synchronized with the radar Tx.
    \textbf{Radio as a Sensor:} The radio unit conducts \emph{bi-static sensing}, capturing signal transmissions toward users. The radio signal is reflected toward the user transceiver via reflectors instead of returning directly to the radio unit.
Leveraging this observation, we develop a collaborative learning framework that combines wireless sensing (radar) and the communication link (radio) to gain a holistic understanding of the environment and user mobility, particularly in the context of beam management. We then develop beam tracking and blockage prediction mechanisms to ensure a reliable link.

\textbf{Acquiring Contextual Information:} A key question is: what is the context that Radar lacks? We define context as the information about the users and environment that radar needs to aid in communication beam management. For active users, this context information includes their angular and distance parameters relative to the base station, while for reflectors, it is their precise location, orientation, and size (Sec~\ref{sec:whatiscontext}). Having this context will ensure that radar can identify the user/reflectors in its view and distinguish them from other irrelevant objects in the environment. The next question is how to acquire this context from radio beam training. The base station radio emits a known preamble across multiple beams, received and utilized by active users to estimate Channel State Information (CSI) across angle and bandwidth. We developed algorithms that leverage this two-dimensional CSI to estimate user and reflector attributes and map them to the radar's view as contextual information as described in Sec~\ref{sec:acquirecontext}. 

% We utilize the communication radio unit to obtain contextual information about active users and environment reflectors.

% This data is extracted during the radio beam scan procedure, where the base station emits a known preamble across multiple beams, received and utilized by active users to estimate Channel State Information (CSI) across angle and bandwidth. We developed algorithms that leverage this CSI to estimate user and reflector attributes and map them to the radar's view as contextual information. 

\textbf{Context-aware Beam Management:} After acquiring context, the radar can distinguish active users and reflectors from other environmental objects. We then developed context-aware beam management by actively tracking the direct and reflected path angle towards all active users for the duration when the radio is busy with data communication. Our tracking solution is inspired by the spatial-temporal consistency of user motion, i.e., a certain maximum speed of motion typically bounds the user location, and further uses Kalman filtering to reduce the impact of noise and provide accurate tracking (Sec~\ref{sec:tracking}). A key challenge, however, is when the direct path is blocked, the radio needs to find a reflected path beam to maintain a reliable connection. To ensure reliability, we track the reflected path angle via all reflectors identified in the previous radio beam training. We estimate the reflector parameters (location, orientation, and size) and develop a geometry transformation to derive the reflected path angle as a function of the user's direct path angle and known reflector parameters. This way, we track the direct and reflected path angles for all time. The base station uses these angle estimates to maintain a reliable beam towards the user via the direct or reflected paths, even in the presence of intermittent blockages. We developed proactive blockage management schemes that detect blockages in advance and instruct the radio to switch the communication beam to one of the unblocked paths to maintain a reliable connection. This is only possible with \name since we actively maintain both direct and reflected path angles using context-aware radar tracking.

% \textbf{Context-aware Proactive Blockage Mitigation:} The final challenge is to detect any blockage event and trigger a counter mechanism proactively. To detect blockage, we define a blockage zone as an area where a blocker could block the link between the base station and the user. We then use the radar measurements to actively look for blockers entering or exiting this region and classify those objects as blockages. Once a blockage is detected, we switch the beam at the base station to one of the unblocked paths to maintain a reliable connection. 

\textbf{Evaluation overview:} We evaluate \name with an experimental setup that consists of a 28 GHz mmWave radio and 24 GHz COTS ISM band radar with four mobile users. Our setup supports mmWave 5G NR waveform for beam scan procedure. We synchronously collect radio beam scan measurements and radar frames for the duration of the experiment and evaluate \name and other baselines on this dataset.
% \footnote{We will make our dataset and code artifacts available to the research community.}. 
The results show that \name maintains a reliable and high throughput link while tracking mobile users and proactively handles blockages using radar measurements while periodically collaborating with the radio sensor. The median throughput is improved by 2.5x, while the worse-case 20th percentile throughput is improved by 8x while keeping the communication overhead as low as 0.5\%.

% \todo{Add Contributions to emphasize on novelty}

%% file: 2_related_v2.tex
% !TEX root = main.tex

\section{Background and Related Work} \label{sec:related}

% Multi-user mmWave communication requires managing directional beams for all mobile users under the impact of reflectors and blockers. 

% \reviewerTwo{No comparison with state-of-the-art and ML papers.} \\
% \reviewerFive{No comparison with past work based on LiDAR or WiFi.}
\textbf{Radar-aided mmWave systems:} 
Radar provides high-resolution sensing of various object's location, distance, and velocity in the environment~\cite{pearce2023multi, ku2023characterizing}. They are low-cost and can operate in ISM bands such as 24 GHz, 60 GHz, or 77 GHz, which does not require spectrum auctioning~\cite{radarbook2}.
FMCW radars have been used for tracking mmWave radio beams using theoretical stochastic geometry models~\cite{nabil2024beamwidth}, and machine learning~\cite{demirhan2022radar,luo2023millimeter,aydogdu2020distributed, chen2023beam,chen2023radar}.  These works exploit an end-end machine learning approach where raw range-angle and range-Doppler images are fed into the ML model to get the beam index. Here, only the direct path angle is tracked using the radar, and a single-user system is assumed, which does not address the complexity of multi-user scenarios with reflectors. \cite{aydogdu2020distributed} relies on GPS at the user for beam prediction, but GPS is unreliable and unavailable on all mmWave devices.
All these works are still primitive in handling multiple devices, lack beam tracking over time and integration with mmWave beam scan procedure, and do not focus on system integration. 
\name is the first end-to-end radar-driven mmWave communication system that leverages collaborative learning, user tracking, reflector modeling, and blockage mitigation to maintain a high throughput and reliable link.

\textbf{Other sensors-aided mmWave systems:} 
LiDARs~\cite{woodford2021spacebeam,klautau2019lidar, jiang2022lidar} and cameras~\cite{alrabeiah2020millimeter, charan2023camera, imran2023environment, ahmad2023vision, charan2023user, nie2023vision} are demonstrated to track directional beams using signal processing and machine learning models. However, these models demand re-training for diverse environments, often involving labor-intensive data collection efforts with ground truth information, yet may still fall short of obtaining the requisite contextual awareness. In fact, the foundation of contextual awareness laid in \name can be applied to these sensing modalities to further improve their performance and generalize them to any indoor or outdoor scenarios.
% However, these sensors are expensive, privacy-intrusive, and non-ubiquitous for NextG cellular networks.  
% Our choice of Radar sensor is inspired by its use of radio-frequency antennas and circuits which can be well integrated with the Radio. 
Other sensor-aided communication using device positioning systems \cite{wei2017pose,va2015beam}, light sensors~\cite{haider2018listeer}, and out-of-band WiFi~\cite{sur2017wifi,nitsche2015steering} are less accurate and are mainly developed for indoor WiFi setting which is not common for NextG cellular deployment scenarios.

% \reviewerTwo{Bold claims - past works assume no clutter, CommRad is not first end-to-end radar-driven mmWave communication systems that leverage collaborative learning, user tracking, reflector modeling, and blockage mitigation tools. CommRad has many simplifications as well and no significant delta is conveyed.}

 % To counter this issue, additional sensors, such as GPS or IMU, can be deployed on the user side to provide user identity to the radar. 
\textbf{High-overhead beam training:} 
There is extensive work in beam training methods to reduce the codebook size (scan fewer beams), but since they have to cover an entire angular space to detect a new user or reflector, ---which can be anywhere in the angular space--- they make a compromise of higher misdetection rates to reduce scanning overhead, especially in multi-user mobility scenarios~\cite{jeong2015random,zhou2012efficient,barati2016initial, sur201560, hassanieh2018fast, aykin2019smartlink, heng2023grid}.
To reduce this overhead, \cite{wang2020demystifying, heng2024site} relies on a pre-determined set of beams that are obtained for a site-specific location (e.g., road or highway) for V2X scenarios, but it does not generalize to arbitrary 6G cellular deployments.

\textbf{Reflector Modeling:}
The authors of \cite{zhou2012mirror,genc2010robust,zhao2018improving,khawaja2018coverage, zhu2014demystifying,wei2017facilitating,zhou2019autonomous, zhou2020robotic, zhou2017beam} leverage reflective surfaces to improve indoor mmWave connectivity \& coverage using only the radio data at multiple static locations. Their model is suitable for one-time modeling during the initial network deployment, which complements our goal of improving the link throughput and reliability in real-time for mobile users using not only radio data but also both radar and radio collaborative learning.

\textbf{Blockage Prediction:} 
The authors of \cite{demirhan2022radarblock} developed radar-aided blockage prediction using an end-end machine learning model. BeamSpy~\cite{sur2016beamspy}, Terra~\cite{ganji2022terra} handle blockage by stationary objects without modeling mobile links and cannot adapt dynamically to blockage arrival patterns. Multi-connectivity~\cite{zhang2018mmchoir} and handovers~\cite{sur2018towards} are alternative ways of recovering from permanent blockages. In contrast to these works, we explore dynamic blockage events where a mobile blocker crosses the link or the mobile user crosses a static/mobile blocking object. Unlike~\cite{demirhan2022radarblock}, our end-end framework not only predicts blockages but also integrates with direct/reflect-path beam tracking with contextual awareness.

%% file: 3_design_v4.tex
\section{\name Design} \label{sec:design}
In this section, we discuss the design and implementation of \name, a sensing-driven mmWave communication system to aid in mmWave beam management. Managing directional beams in mobile multi-user scenarios requires maintaining the beam direction towards the user either via a direct path or via a reflected path when the direct path is not available due to blockage events. The key information needed for this beam management at the radio is a time series of the best beam angles for all users at all times. The radio can obtain this information through the periodic beam training phase, but this information gets outdated when the radio is not doing beam training, i.e., it is communicating with active users. During this time, the radio relies on the radar sensor to manage its beams. The radar sensor is great for detecting objects in the environment with high resolution in the distance and angle estimates thanks to the high bandwidth available (over GHz) in free ISM bands such as 24 GHz, 60 GHz, or 77 GHz center frequencies. Therefore, in \name, we explore integrating a radar sensor with the radio to aid beam management. To aid in radio beam management, the radar sensor must provide the time series of the best beam angles (direct or reflected path angles) for each user at all times. However, the radar sensor receives an extensive amount of reflections from many objects in the surroundings, and it cannot distinguish which objects are relevant for beam management, such as users, reflectors, and blockers. 

\begin{figure}[t!]
    \centering
    \includegraphics[width=.47\textwidth]{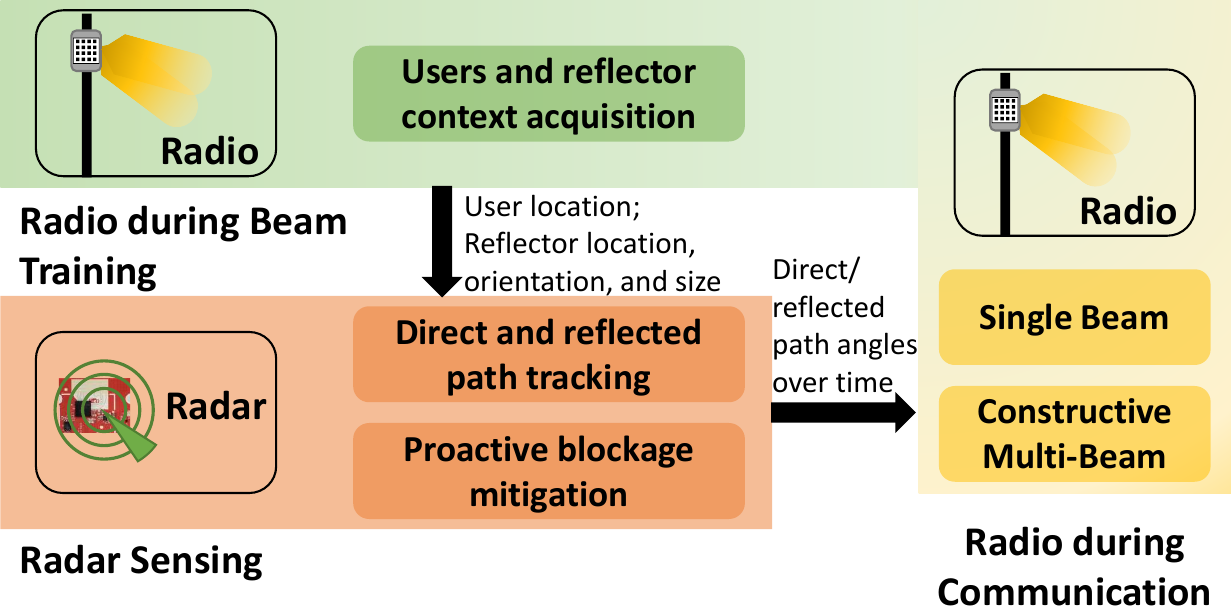}
    \caption{Overview of \name's end-end system implementation of collaborative bi-directional learning with radar+radio integration.}
    \label{fig:overview}
\end{figure}

In \name, we developed a three-stage process to acquire this crucial context and build a bi-directional collaborative framework for beam management, an overview of which is shown in Figure~\ref{fig:overview}. \textbf{(Stage 1) What is context?} The first step is to define the context for the radar so that it can aid in beam management. The context should be relevant for identifying users, reflectors, and blockers in the environment and should be understandable by the radar, i.e., mapped into the radar's range-angle view. We discuss this context and how it is hard to obtain by the radar in Section~\ref{sec:whatiscontext}.
\textbf{(Stage 2) How to acquire context?} The second stage is to acquire this context for the radar. More importantly, the context of the users and reflectors should be updated periodically since the radar may lose this context in a dynamic environment where new users may appear or existing users may disappear. We discuss our algorithms for acquiring this context through the periodic radio beam training process in Section~\ref{sec:acquirecontext}. \textbf{(Stage 3) How to do context-aware beam management?} Finally, as the radar is equipped with context, it has to continuously aid mmWave radio in maintaining the beams when the beam training is not immediately available. We discuss how radar aids beam management by tracking the user's direct and reflected path and proactively detecting blockers in Section~\ref{sec:tracking}. Actively tracking the reflected path along with the direct path enables us to implement advanced constructive multi-beam techniques~\cite{rappaport2013millimeter,jain2021two} to improve link reliability. The overall system is implemented through a synchronized COTS radar and radio platform and evaluated in real-world settings, as discussed in upcoming sections.

\subsection{What is the Context that Radar Lacks?}\label{sec:whatiscontext}
The radar sensor detects different objects in the environment in the form of their distance and angle estimates. It creates a distance-angle view that shows multiple points of reflection from objects with varying strengths. We define the context for the radar as the set of points that belong to the relevant objects for the communication radio, such as users, reflectors, and blockers. Other objects in the environment that do not fall into this category are not relevant to the scope of this paper. \textbf{(1) Context for users:} For the users, the context lies in the points in the distance-angle view that belong to an active user for which the radio has to maintain a directional beam for communication. Here, we define context as the approximate \{distance, angle\} of points that belong to an active user. \textbf{(2) Context for reflectors:} For the reflectors, the location is crucial too, but not enough. A reflector is modeled as a line segment, which is characterized by the location (any point on the reflector), orientation (slope of the line segment), and size (the start and end points on the reflectors). Learning these reflector parameters is essential to compute the reflected-path beam angle. \textbf{(3) Context for blockers:} Finally, we define the context for the blockers by their relative location, size, and speed, which is necessary to estimate the arrival time and duration of blockage events.
Without this context information, the radar sensor cannot aid mmWave radio in direct/reflected path beam maintenance.

\subsubsection{\textbf{How Radar Lacks Context for Users?}}
\begin{figure}[t!]
    \centering
    \includegraphics[width=0.5\textwidth]{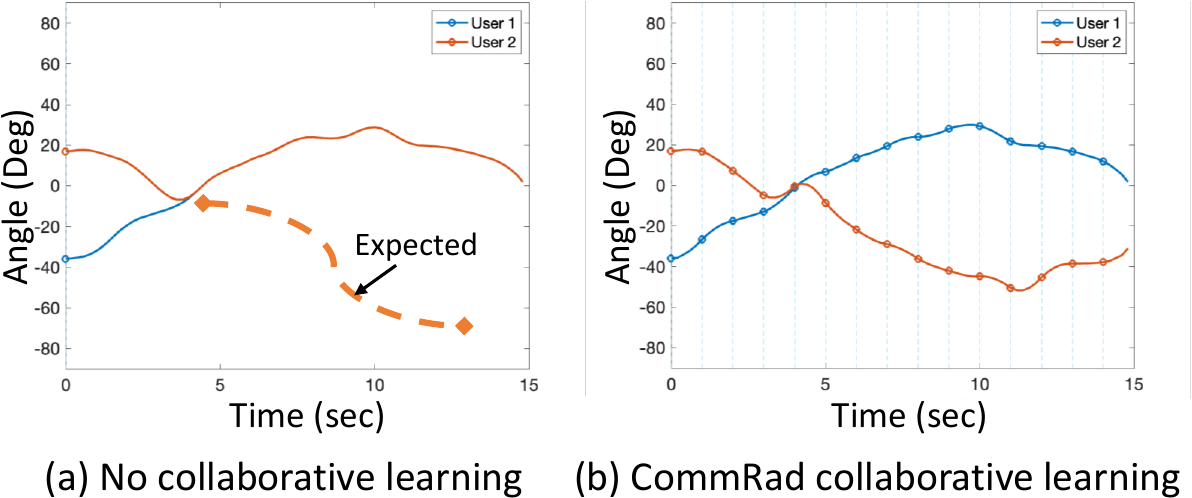}
    \caption{Challenges with lack of context for user identification}
    \label{fig:with_collaborative_learning_tracking}
\end{figure}

~\\
The radar sensor detects multiple objects in the environment, but it does not know which objects belong to an active mmWave user. Without this crucial context, Radar cannot track user mobility and aid in beam maintenance. To demonstrate this limitation, consider a scenario with two mobile users in the environment crossing each other, i.e., one user is moving left to right, and the other is moving right to left in front of the radar. The radar can detect two users and track their angle over time. Without context information, the radar cannot distinguish between the two users 
and incorrectly assumes that both users continue along the same trajectory after the crossing, as shown in Figure~\ref{fig:with_collaborative_learning_tracking}(a). In reality, they move in opposite directions. This confusion can be resolved by periodically incorporating contextual information from the radio sensor, resulting in accurate mapping of user trajectories to user ID, as demonstrated in Figure~\ref{fig:with_collaborative_learning_tracking}(b).

% The radar sensor offers precise location data for various objects in the environment Within this array of tracked objects, the primary challenge arises in discerning which ones qualify as active users deserving of a dedicated beam from the base station. The radar does not have context of what a point in the point cloud corresponds to. Radars primarily capture specular reflections only, i.e., they have a sparse set of points from each object, rendering user identification a complex task. 

% Furthermore, even if we perform a one-time reflection on user mapping, that won't be sufficient since user mobility can lead to identity loss over time.  A straightforward approach is to base this association on motion constraints, assuming that users can only move within certain distances over defined time intervals, given speed constraints. However, this method falls short in scenarios such as users crossing each other as illustratedin Figure~\ref{fig:with_collaborative_learning_tracking}(a). Despite having their initial identities (mapping of radar reflection to radio-users), the radar loses track of User1's identity when the two users cross paths and incorrectly assumes that both users continue along the same trajectory after the crossing, while in reality, they move in opposite directions. This confusion can be resolved by periodically incorporating contextual information from the radio sensor, resulting in accurate tracking of all users, as demonstrated in Figure~\ref{fig:with_collaborative_learning_tracking}(b).

\subsubsection{\textbf{How Radar Lacks Context for Reflectors?}}
\begin{figure}[t!]
     \centering
     \begin{subfigure}[b]{0.49\linewidth}
        \centering
        \includegraphics[width=\linewidth]{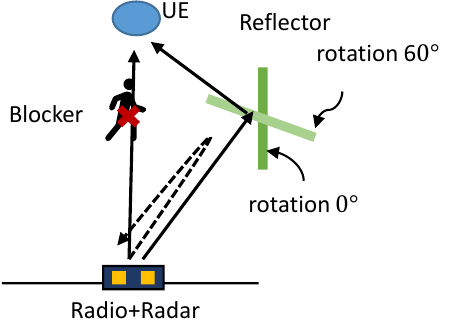}
        \caption{Varying reflector orientation experiment setup}        
     \end{subfigure}
     \hfill
     \begin{subfigure}[b]{0.49\linewidth}
        \centering
        \includegraphics[width=\linewidth]{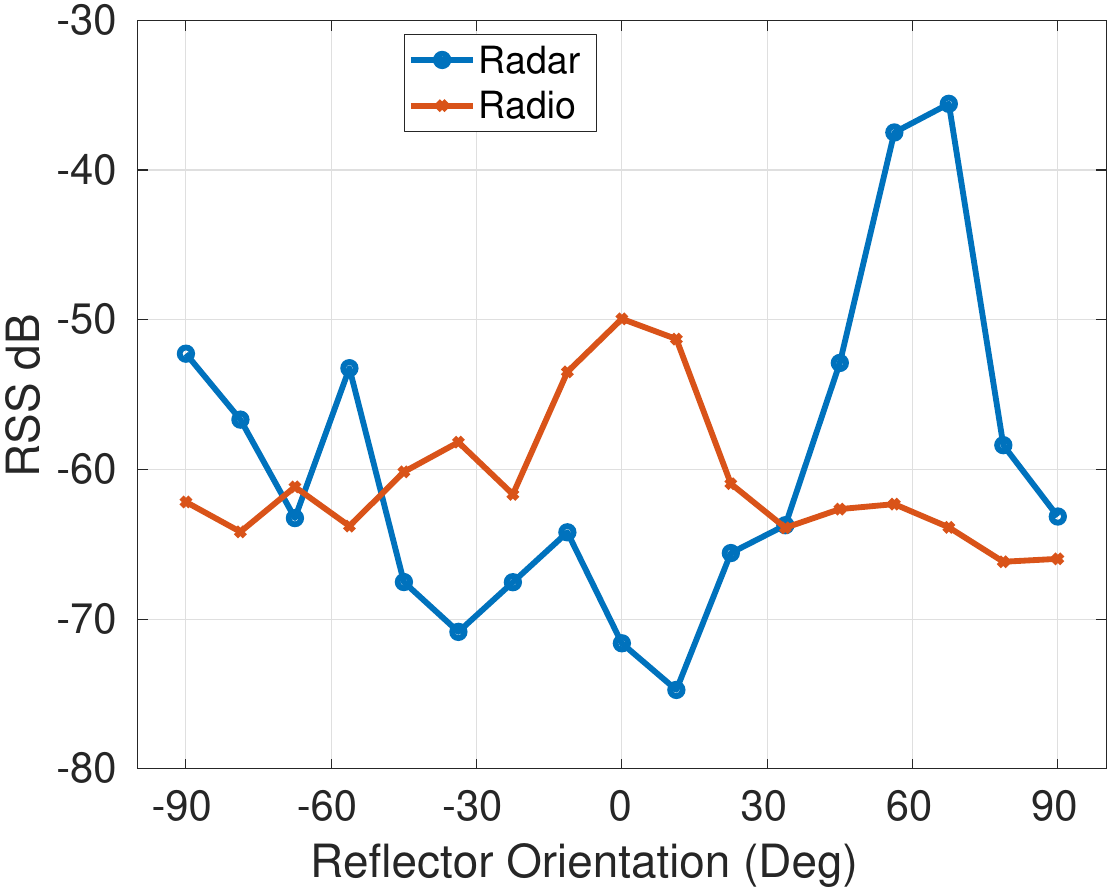}
        \caption{RSS with varying reflector orientation}
     \end{subfigure}
        \caption{Challenges with lack of context for reflector identification.}
        \label{fig:design:refl_rotation}
\end{figure}

~\\
% Reflectors help in maintaining a reflected-path link to the user when the direct path link is not available. So, how can we use radar to learn a reflected-path link? 
% The next challenge lies in identifying useful reflectors, objects in the environment that can redirect the signal toward the user device when the direct path link between the base station and the user is obstructed. 
The next challenge for radar is determining which objects in the radar's view serve as useful reflectors. Simply gauging signal strength on the radar to locate strong reflectors is not feasible, as objects that appear as a strong reflector for radar could be weak for communication. To illustrate this, we conducted an over-the-air experiment with our testbed involving varied reflector orientations. We plotted the Received Signal Strength (RSS) at both the radar and radio, as depicted in Figure~\ref{fig:design:refl_rotation}(a). When the reflector is positioned at 0 degrees orientation, it optimally serves the user with the highest RSS, whereas for other reflector orientations, RSS diminishes as shown in Figure~\ref{fig:design:refl_rotation}(b). In contrast, the radar obtains its highest RSS when the reflector is oriented at 60 degrees, a configuration that does not effectively serve the user through the reflected path, adhering to the law of reflection. We show in later sub-sections how radar-radio collaboration helps in obtaining the context for the reflectors and aiding in reflected beam maintenance even under user mobility.


% \todo{How radar tracks each user while collaborating with the radio}
\subsection{Acquiring Context through Radar+Radio Collaboration}\label{sec:acquirecontext}
So far, we have seen that the radar sensor alone cannot obtain contextual information about users and reflectors independently. This section describes how contextual information can be obtained using our radar+radio framework. This contextual information is then used to provide context to radar to map different objects as users or reflectors.

\subsubsection{\textbf{Acquiring Context for Users} }
\begin{figure}[t!]
    \centering
    \includegraphics[width=.5\textwidth]{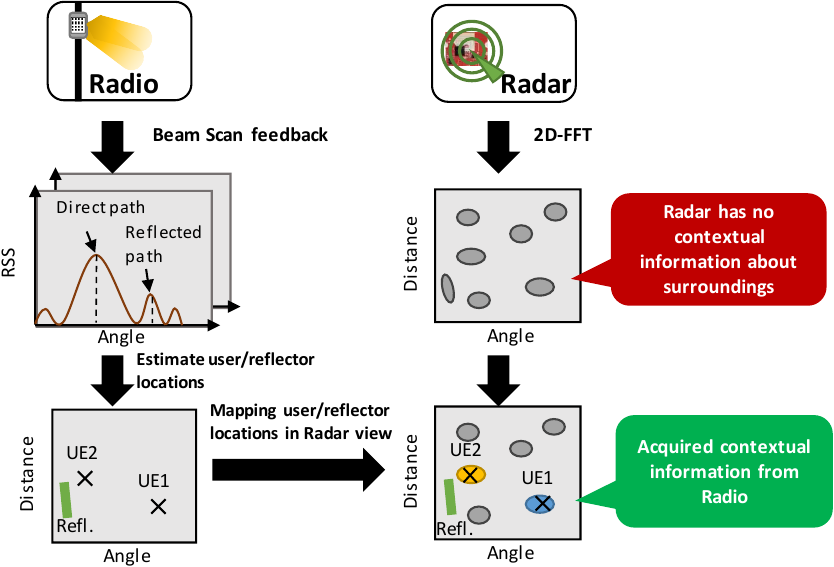}
    \caption{Identifying users and reflectors in radar view using contextual information from radio.
    % \reviewerOne{Replace MUSIC with 2D-FFT
    }
    \label{fig:user_location}
\end{figure}

~\\
% \todo{this is really short sub-section... maybe combine with the next one??... Acquiring beam-management context from radars.. }
Acquiring user context is a process of identifying which points in the radar's distance-angle view are active users. We use the radio beam scan procedure to estimate the user's angle and distance from the base station and then map it to the radar's view. First, consider a simple multi-user scenario without any reflectors and blockages and later analyze their impact. During the beam scan process, the base station transmits the preamble through multiple beams in different directions. Each user captures this signal and estimates the RSS per beam. The beam index that gives the highest RSS is identified as the best beam that serves the individual user. The direction at which this beam points is identified as the user's angle. We further estimate the user distance using the observation that the distance is inversely proportional to the RSS using Friss equation \cite{shaw2013radiometry}. These estimates are used to map the user's location in radar view as shown in Figure~\ref{fig:user_location}.

\subsubsection{\textbf{Acquiring Context for Reflectors} }~\\%identification in \name
Reflector identification is a crucial process in \name, as reflectors are key in maintaining reliable communication~\cite{rappaport2013millimeter,jain2021two}. With its monostatic sensing capabilities, radar faces challenges in independently identifying reflectors, which contrasts with the bistatic nature of communication systems. In \name, we leverage distinctive attributes from both radar and radio to acquire reflector identification.

Reflectors can be categorized as point reflectors (e.g., poles and lamp posts) or surface reflectors (e.g., buildings, glass walls, and monitor screens). Surface reflectors are particularly relevant for mobile links since their reflections are well-defined and predictable, unlike point reflectors, which are better suited for static users as they struggle to sustain mobile links.
% , as depicted in Figure~\ref{fig:scatter_vs_reflector}. 
Therefore, our focus in this paper is primarily focused on surface reflectors.

We represent surface reflectors as line segments in Cartesian coordinates, characterized by their location (any point on the reflector), orientation (slope of the line), and size (start and end points on the reflector). We estimate these parameters leveraging two-dimensional channel measurements across angle and bandwidth, following a three-step process as follows:

% To estimate these reflector parameters, we employ the radio beam scan procedure. Leveraging 
% radio beam scan Channel State Information (CSI) from each user across both beam scan angles and bandwidth, as illustrated in Figure~\ref{fig:reflector_estimation}, we follow a three-step estimation process:

\begin{figure}[t!]
    \centering
    \includegraphics[width=.5\textwidth]{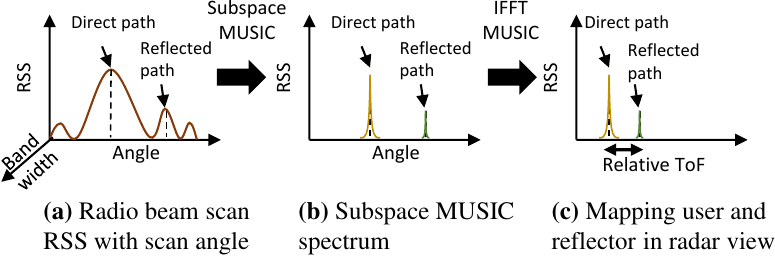}
    \caption{Reflector location estimation using radio beam scan measurement.}
    \label{fig:reflector_estimation}
\end{figure}

\textbf{(i) Estimating reflected path angle and distance:} As a first step, we must separate the direct path and reflected paths in the radio RSS measurements. We apply a variation of the MUSIC algorithm, ``Subspace MUSIC," to separate the two paths across the angle domain, as we do not have per-antenna measurements in mmWave systems with analog beamforming arrays. We then apply MUSIC across bandwidth to obtain relative Time of Flight (ToF) information between the two paths, following the classic procedure in localization~\cite{kotaru2015spotfi}. The direct path is identified as the one with the lower ToF, corresponding to the shorter signal travel distance. By utilizing the relative ToF and actual user location obtained through RSS measurements, we determine the angles and distances of both the direct and reflected paths. The procedure is explained in Figure \ref{fig:reflector_estimation}.

\begin{figure}[t!]
    \centering
    \includegraphics[width=.45\textwidth]{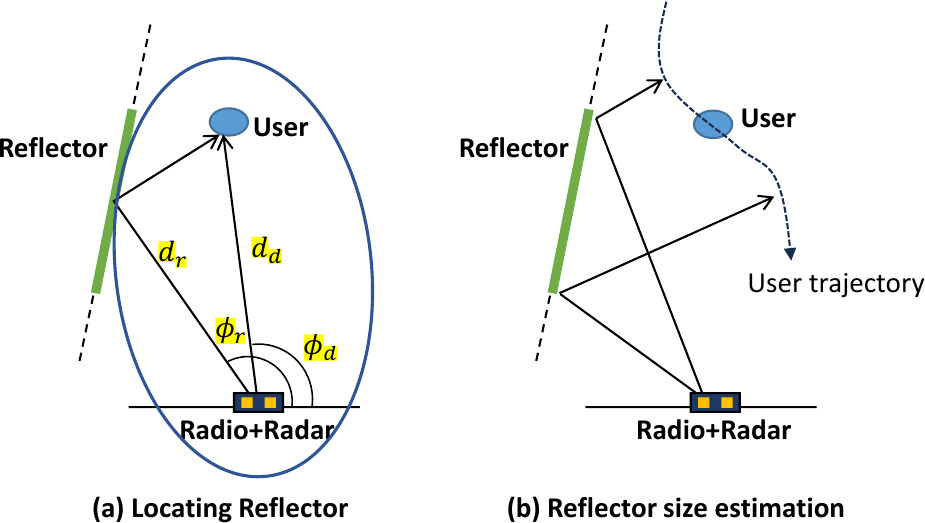}
    \caption{Reflector orientation and size estimation.}
    \label{fig:reflector_estimation2}
\end{figure}

\textbf{(ii) Estimating reflector orientation:} After obtaining the angles and distances of each path, we construct an ellipse with the base station and user location as its foci. The reflector's position locus lies on this ellipse, as shown in Figure~\ref{fig:reflector_estimation2}(a). To pinpoint the exact reflection point on the ellipse, we find the intersection between the ellipse and the line extending from the base station towards the reflector angle. This intersection point serves to identify the orientation of the reflector as the slope of the tangent line to the ellipse at that location.

\textbf{(iii) Estimating reflector size:} To complete the reflector characterization, we need to estimate its size, which is crucial for tracking the reflected path angle. We address this challenge through a learning process over time, as demonstrated in Figure~\ref{fig:reflector_estimation2}(b). We collect multiple beam scan measurements from various user locations as they move around and repeat steps (i) and (ii) to identify all the points on the reflectors. In this way, we obtain the reflector size as a set of two endpoints on the line segment obtained in this process. Notably, this is a one-time process for static reflectors but can be repeated if changes occur in the reflector configuration.

% \begin{figure}[t!]
%     \centering
%     \includegraphics[width=0.45\textwidth]{figures/scatter_vs_reflector.pdf}
%     \caption{Comparison of scattering random reflections vs. structured reflector reflections.}
%     \label{fig:scatter_vs_reflector}
% \end{figure}

% \subsubsection{Reflector identification using radio}

% \textbf{radio-Radar Collaborative learning to enhance reflector identification:} 
\textbf{Problem of unreliable RSS measurements:} It is essential to note that the reflector identification process may accumulate errors if underlying parameters are inaccurate. A key source of error is distance estimation from RSS measurements, especially in multipath environments. We have observed that radar distance measurements are more accurate, relying on time-of-flight measurements from a synchronized Tx and Rx architecture. Thus, our approach initially obtains course distance estimates from radio measurements and then refines them by identifying the nearest object in the radar view. This dual-step \textit{radar-radio collaborative learning} process ensures accurate distance and angle measurements, enhancing reflector identification accuracy.

\subsection{Context-Aware Beam Management using Radar Sensor}\label{sec:tracking}
After identifying users and reflectors in the radar view through radar-radio collaborative learning, we next use the radar sensor to track the user motion through changes in both the direct path and reflected path over time. It is important for radar sensors to maintain and aid direction beams while the radio is occupied with data communication and not performing beam scans. Our approach to the direct path and reflected path tracking is described in this subsection.

Before discussing our tracking algorithm, we present a primer on the choices of range and resolution, followed by the techniques used to counter the challenges described in the previous sections.

\textbf{Maximum Range and Range-Resolution:} \\ 
The range resolution for FMCW radar is given by $\Delta \text{R} = \frac{c_0}{2B}$. CommRad uses a bandwidth of 200MHz, thus resulting in a range resolution of 0.75m. 
With a resolution under 1m, we are able to capture the movement of the user across time accurately. 
The maximum range of the FMCW radar is given by $
\text{R}_{max} = \frac{c_0}{2B} N$,
where N is the number of samples corresponding to the duration observed. For an N value of 1000, we get a maximum range of 7.5km. 

\textbf{Angle resolution:}
The angle resolution of a antenna array with $N_{ant}$ is given by, $\theta_{res} = \frac{2}{N_{ant}}$. For a 16 virtual antenna system, angular resolution is $7.16^o$, whereas for a 8 antenna system, angular resolution is $14.32^o.$
By using a virtual antenna array, we achieve better angular resolution, which helps improve the localization and tracking accuracy of the system.

\begin{figure}
    \centering
    \includegraphics[width=0.5\textwidth]{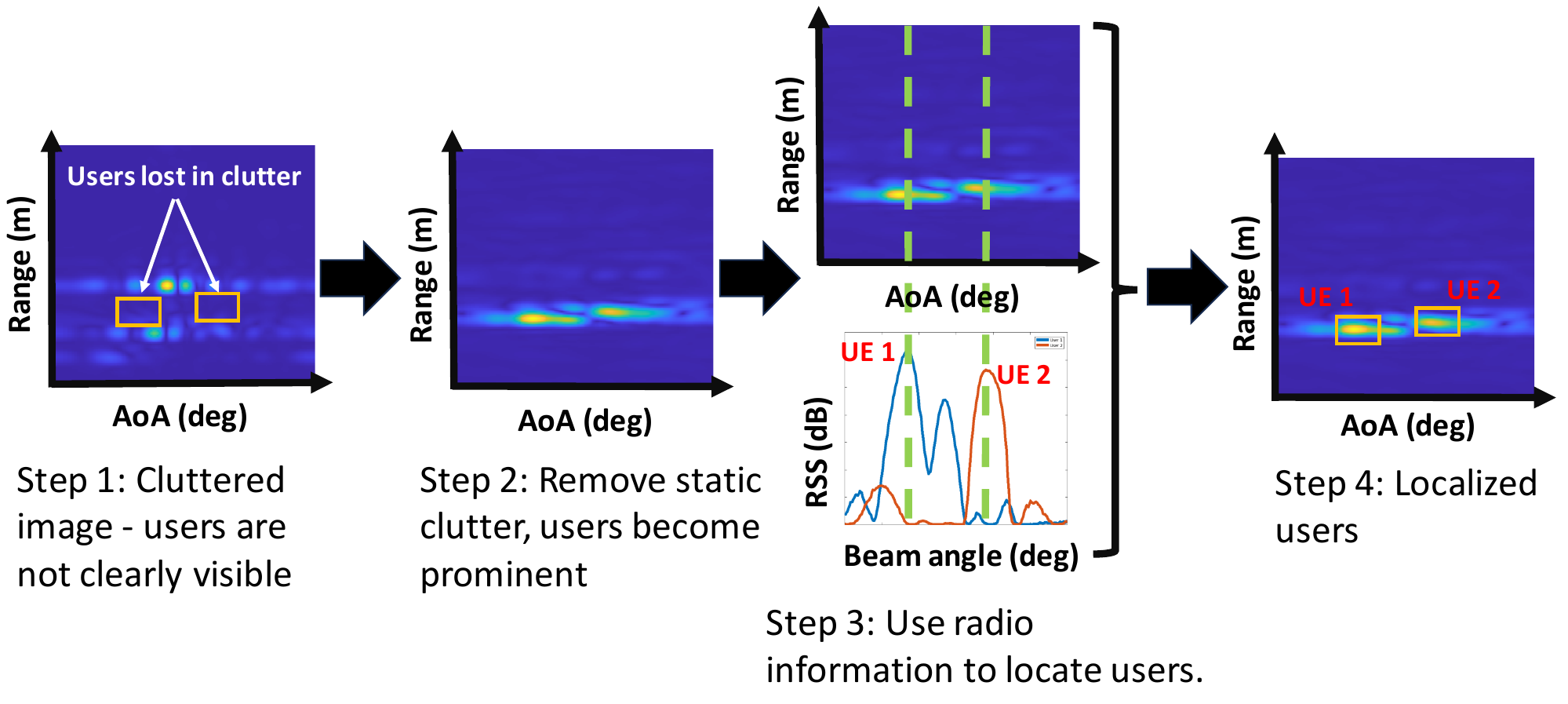}
    \caption{Clutter removal and user localization}
    \label{fig:clutter_removal}
\end{figure}
\subsubsection{\textbf{Static Clutter removal with frame subtraction}}
The Range-AoA representation from radar contains high-energy points corresponding to all environmental reflections, including users, reflectors, blockers, etc. While tracking the user, all static reflections coming from the environment can be considered as clutter. The presence of clutter causes ambiguity in distinguishing the users over time. To remove clutter, we use the static nature of those objects and cancel them across time. Each frame, corresponding to one chirp, is 1ms apart from the consecutive frame. Depending on the type of mobility (users in indoor scenarios have lesser mobility than in outdoor vehicular scenarios), we perform an average over multiple chirps and subtract the Range-AoA profile of the current frame from the averaged Range-AoA profile. This removes the static objects, and user tracking can be done accurately and reliably. 
We show the above-mentioned workflow in Figure \ref{fig:clutter_removal}. We first obtain the clutter-removed image using frame subtraction and then use the radio data to localize the users within this image.

\subsubsection{\textbf{Direct path beam tracking}}
\begin{figure}
    \centering
    \includegraphics[width=.5\textwidth]{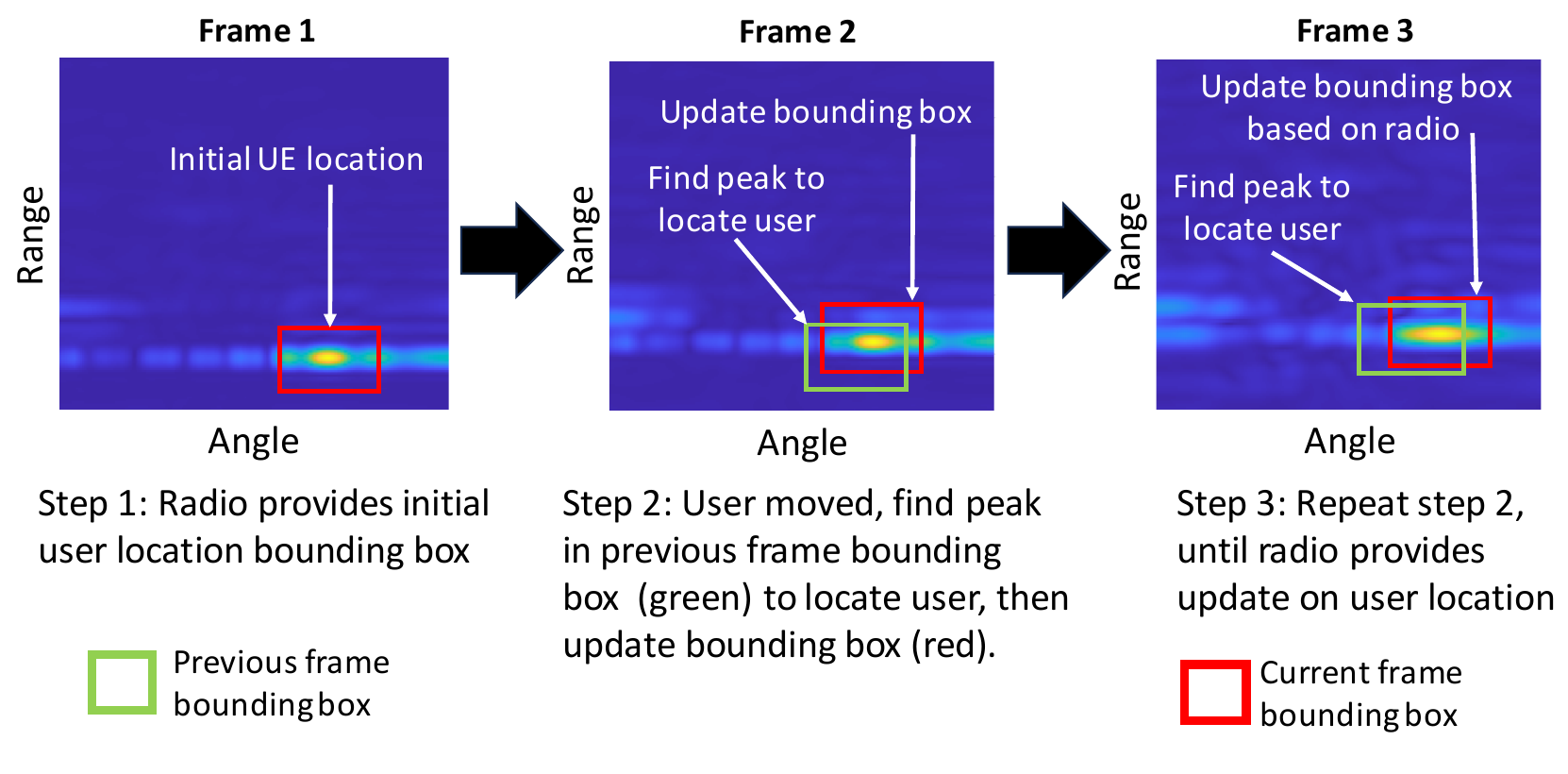}
    \caption{Proposed user tracking algorithm.}
    \label{fig:user_tracking}
\end{figure}

~\\
For direct path tracking using the radar sensor, we have devised a three-step process illustrated in Figure~\ref{fig:user_tracking}. The first step involves identifying the user's initial location within the radar view and creating a bounding box around each user.  Creating a local bounding box around a user allows us to remove the dynamic clutter in the environment, like other users, blockers, etc., that may not be present near the user. In the second step, we examine the subsequent radar frame to locate the user with the highest signal strength while retaining the same bounding box as in the previous step. Finally, in step 3, we adjust the bounding box to align with the new user location identified in step 2. These last two steps are repeated for all consecutive frames to ensure continuous user tracking. By tracking the user with radar, we can still maintain directional beams toward the user, even when the radio resources are unavailable for beam training. Periodic updates of user location from the radio can then help maintain the tracking accuracy of the radar, especially when the radar loses the context of the users who might be too close.  \\

% Once we have the users rough angle and distance information, we map this into radar view frame and identify corresponding user location in the radar view. Figure~\ref{fig:user_tracking} describes user tracking in three steps. The first step is to identify user location in radar view and create a bounding box around each user. The second step for tracking is to look into the next radar frame and locate the user with the highest signal strength using the same window. In step 3, we upgrade the window box to align with the new user location found in setp2 and repeat these two steps for all frames for tracking. 

% \todo{describe the process of mobility tracking. Give range and resolution analysis of radar. How the data is represented as x-y images and how we detect peaks to identify user motion and track over time. Kalman filter model and analysis. Challenges with multi-user scenario and how we address those challenges. Any super-resolution ideas or algorithms given additional context from the radio. The final output - angles of all users (direct path and reflected path)}

% \textbf{Adapting radio direct path beam angle:}

\begin{figure}
    \centering
    \includegraphics[width=.4\textwidth]{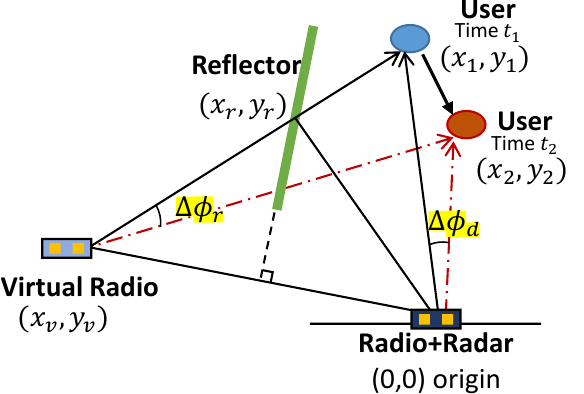}
    \caption{Tracking both direct path and reflected path.}
    \label{fig:tracking_both}
\end{figure}

% Tracking the reflected path is notably more intricate than tracking the direct path. While the angle of the direct path can be directly monitored using radar, tracking the angle of the reflected path necessitates a thorough understanding of the reflector's location, orientation, and size. Once this information is available, we can create a virtual representation of the base station relative to the reflector as an object behind a mirror. The line connecting this virtual base station to the user defines the angle of the reflected path. By integrating this approach with user location tracking, as discussed in the previous subsection, we can readily translate changes in user location into adjustments in the reflected path angle.

% We describe this geometric transformation through a mathematical expression.
\subsubsection{\textbf{Reflected path beam tracking}}~\\
Reflector tracking is more complicated than user tracking, as mobile users change the angle of the reflected path. We need to know the reflector's location, size, and orientation to track the reflector. Once this information is available, we can create a virtual representation of the base station relative to the reflector as an object behind a mirror.  The line connecting this virtual base station to the user defines the angle of the reflected path.
Let the user's location at time $t_1$ be denoted as $(x_1, y_1)$, and at a subsequent time $t_2$, the user's location changes to $(x_2, y_2)$ (See Figure~\ref{fig:tracking_both}). Assuming the radar and radio system is located at the origin $(0,0)$ and has a virtual location of $(x_v, y_v)$, we can estimate changes in the reflected path angle, denoted as $\Delta \phi_r$, as follows:

\begin{equation}
\Delta \phi_r = \tan^{-1}\frac{y_2-y_v}{x_2-x_v} - \tan^{-1}\frac{y_1-y_v}{x_1-x_v}
\end{equation}

Here, the virtual location of the radar/radio system $\x_v,y_v$ is unknown, which is dependent on the reflector's location and orientation, defined as:

\begin{equation}
(x_v, y_v) = (\tan(\phi), -1) \times \frac{2(y_r-\tan(\phi)x_r)}{1+\tan(\phi)^2}
\end{equation}

Where $(x_r, y_r)$ represents an arbitrary point on the reflector, and $\phi$ denotes the reflector's orientation angle.

 \begin{figure}
    \centering
    \includegraphics[width=.4\textwidth]{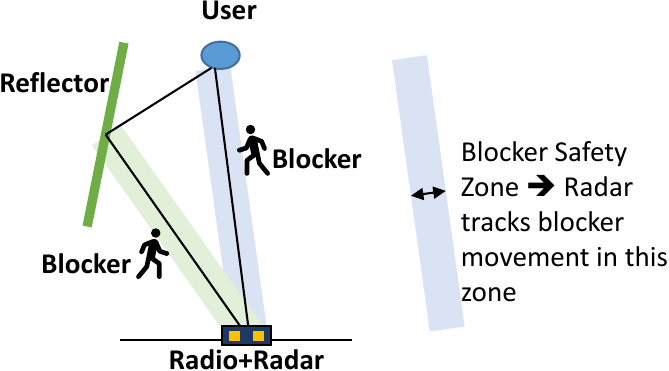}
    \caption{Blockage tracking and proactive prevention.}
    \label{fig:blockage_tracking}
\end{figure}
With the estimate of the reflector orientation, we can track the change in reflected path angle $\Delta \phi_r$, which can be used as an alternate beam path to aid communication when the direct path is blocked or unavailable for communication.
% \textbf{Tracking reflected beam angle:}
% Once we obtain important reflector parameters, then tracking the reflected path angle is a corollary of tracking the user trajectory. Lets say the user moved from location

\subsubsection{\textbf{Context-Aware Blockage Mitigation using Radar}}
Blockage causes trouble to the mmWave link by reducing the signal strength and bringing the link to an outage. Therefore, to revive the link under blockage, alternate reflected paths that are not blocked are utilized. When all the paths are blocked, the link is unrecoverable. To recover the link for partial blockage, it is important to learn the blockage arrival and departure process promptly. Prior work on blockage mitigation is either reactive in nature, which acts when the link is already down, or they rely on radio alone, compromising accuracy in a large multi-user environment.   Our proposed collaborative learning framework takes care of the blockage events. By leveraging unique features of radio and radar sensing, we create a signal processing pipeline to promptly detect blockage and take precautionary actions such as switching beams to a reflected path. 

% -----------**************--------

\textbf{Defining a blockage-prone region:} The first step in detecting blockage with the help of radar is to determine a blockage-prone region. We define a blockage-prone region as a rectangle patch between the user and the base station, as shown in Figure~\ref{fig:blockage_tracking}. Any blocker in this region can cause blockage.
\noindent
The radar tracks an object's velocity over frames to estimate the time-of-arrival of the blocker into the blockage-prone region. The blockage duration is estimated from blocker velocity and blocker size. A blocker of length $l_B$ and velocity $v_B$ would block the link for the duration of $l_Bv_B$.

\textbf{Proactive blockage mitigation:} With the blocker velocity and duration known, the base station alerts the radio of the upcoming blockage event. The response to the blockage depends on the duration of the blockage event. If the blockage event is too slow and lasts longer, then the base station switches to one of the reflected beams. The reflected beam is served until the blockage event lasts. After the blockage is gone, the base station switches back to the original link. 
% \todo{Track any potential blockage for both direct and reflected path}
% \todo{Since we have the information about current user directions, we can figure out if any potential blocker is going to block it. Objective is to find blocker arrival time and blocker size and blocker velocity -> from which we compute blockage duration. How we estimate these blockage attributes? Describe them here.}
 \begin{figure}
    \centering
    \includegraphics[width=.4\textwidth]{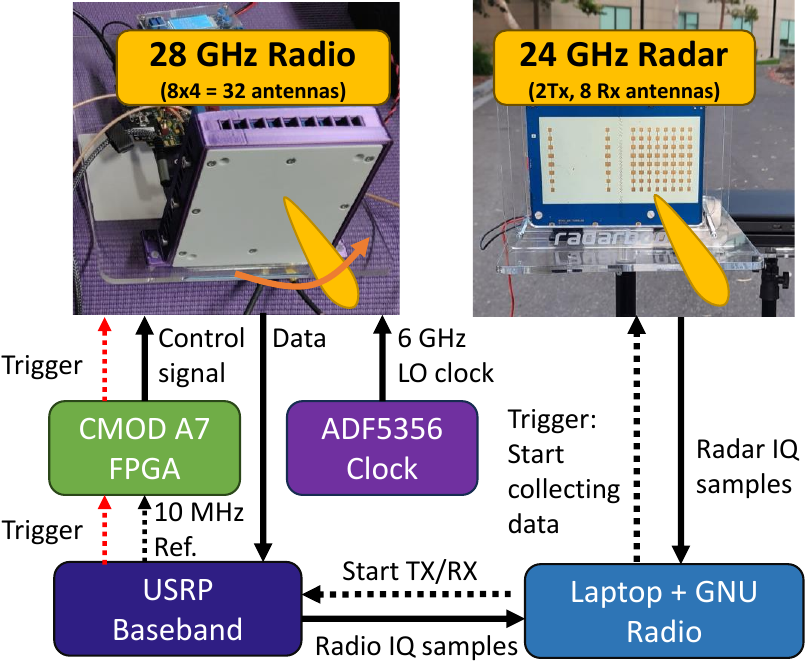}
    \caption{Base station with synchronized Radio and Radar setup}
    \label{fig:exp_setup}
\end{figure}

% \todo{put all together for single beam or CMB management}
\begin{figure*}[t!]
     \centering
     \begin{subfigure}[b]{0.3\textwidth}
        \centering
        \includegraphics[width=\linewidth]{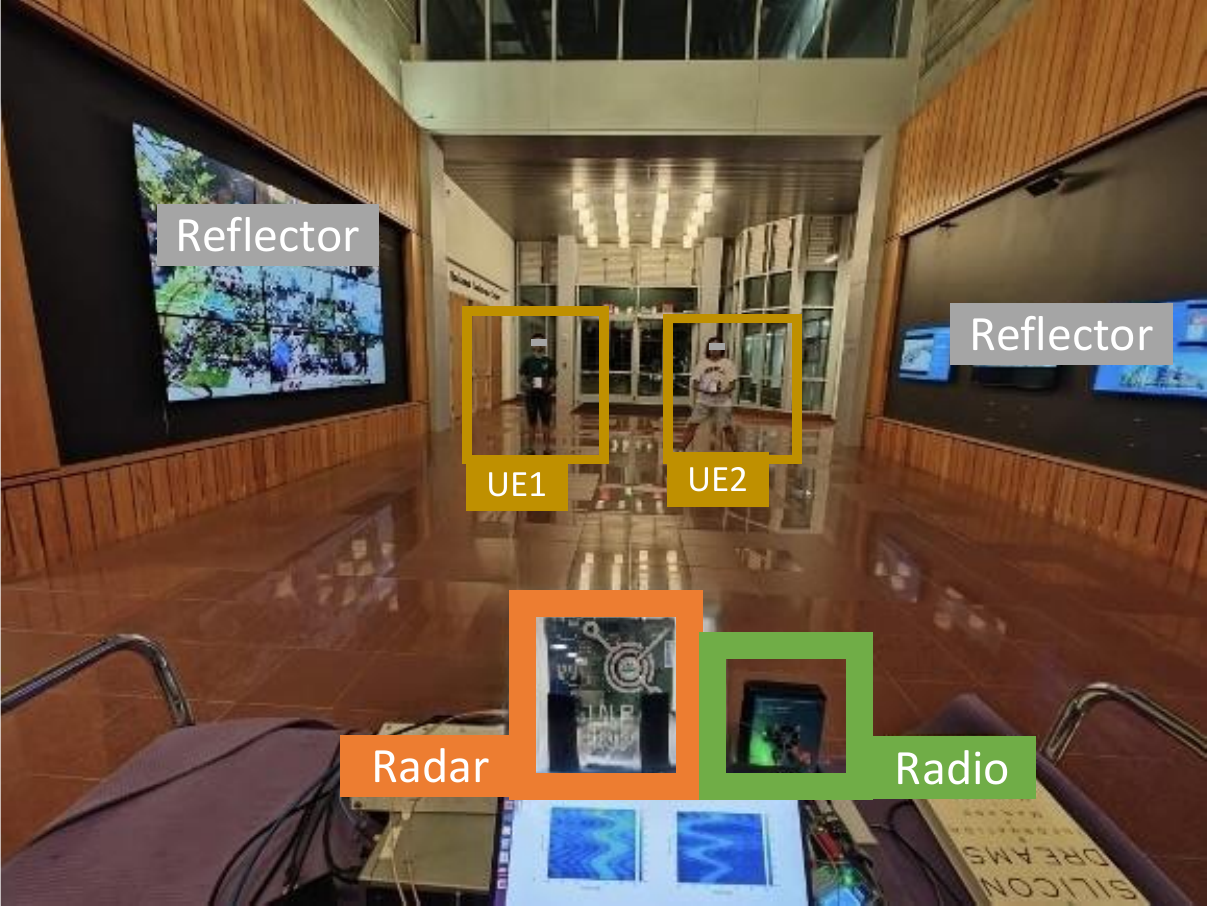}
        \caption{Indoor scenario with reflectors}        
        \label{fig:experiment_indoor}
     \end{subfigure}
     \hfill
     \begin{subfigure}[b]{0.58\textwidth}
        \centering
        \includegraphics[width=\linewidth]{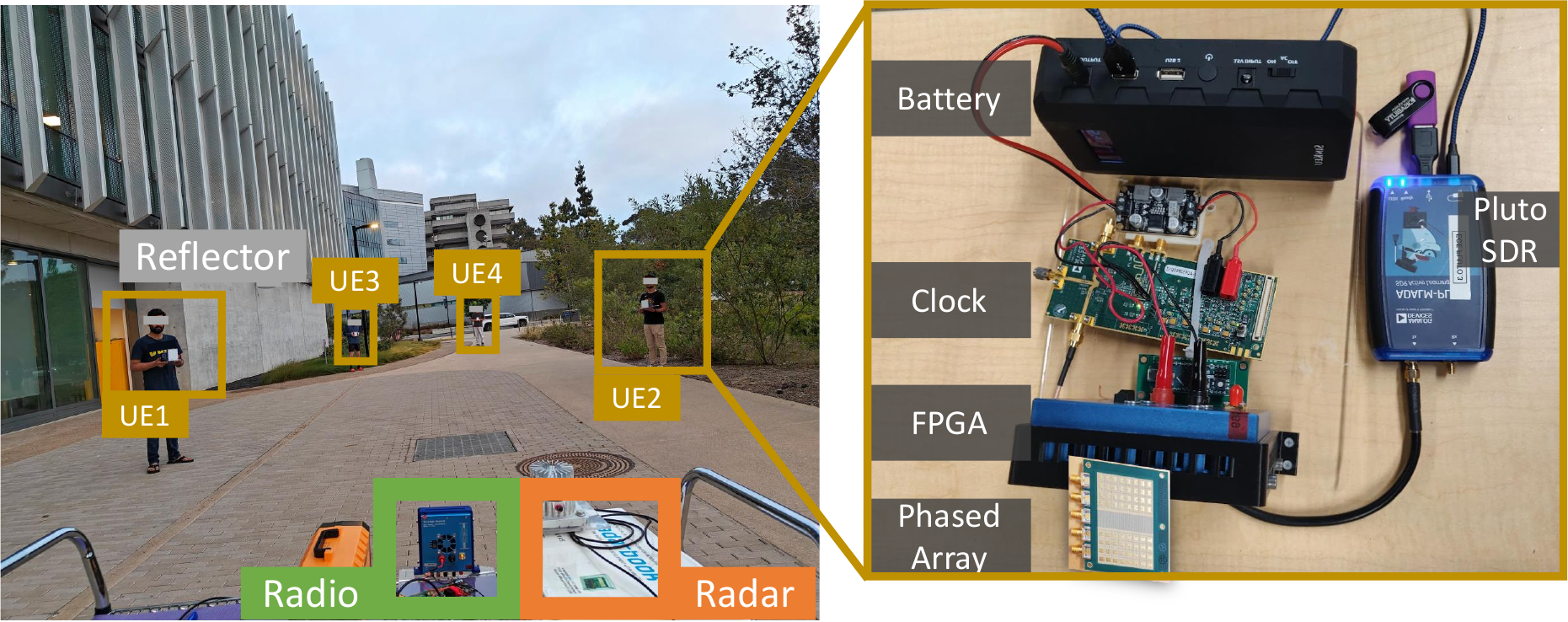}
        \caption{Outdoor scenario with reflectors, four users, and detailed  user setup}
        \label{fig:experiment_outdoor}        
     \end{subfigure}
     % \hfill
     % \begin{subfigure}[b]{0.27\textwidth}
     %    \centering
     %    \includegraphics[width=\textwidth]{figures/demo_setup_ue.pdf}
     %    \caption{User with compact and mobile setup}
     %    \label{fig:2}        
     % \end{subfigure}
        \caption{Experiment scenarios with the static radar-radio platform with four mobile users. We consider diverse scenarios indoors and outdoors with various reflectors such as concrete walls, monitor screens, glass walls, etc.}
        \label{fig:expt_scenario}
\end{figure*}

%% file: 4_implementation.tex
% !TEX root = main.tex
\section{Implementation}
The implementation of the entire system is made in a way that enables experiments that can be conducted both indoors and outdoors. We describe the hardware and software used, the implementation of each subsystem, and the integration of these subsystems to create the joint synchronized platform as shown in Figure~\ref{fig:exp_setup}. 

% \begin{figure}[t!]
%     \centering
%     \includegraphics[width=.5\textwidth]{figures/expt_setup.pdf}
%     \caption{\todo{New figure to show detailed setup and synchronization}}
%     \label{fig:expt_setup}
% \end{figure}

% \begin{figure*}[t!]
%      \centering
%      \begin{subfigure}[b]{0.45\linewidth}
%         \centering
%         \includegraphics[width=\linewidth]{figures/gnb_setup.pdf}
%         \caption{Base station with synchronized Radio and Radar setup}        
%         \label{fig:1}
%      \end{subfigure}
%         \caption{Our mmWave testbed for experimental evaluation }
%         \label{fig:expt_setup}
% \end{figure*}

\subsection{Radio Hardware and Software}
The mmWave radio system is powered by a combination of commercial software-defined radios (SDRs) and custom-designed phased arrays. 
The SDRs are connected to the PC using SFP optical cables or USB cables. UHD and GNU-Radio are used to control the SDRs. 
% The phased arrays are controlled through an FPGA with a Serial Peripheral Interface (SPI) interface. The FPGA is connected to the PC to send the beam parameters. 

\textbf{$\blacksquare$ Phase Array specifications and control hardware:} The phased arrays have 32 antennas in a 4x8 configuration~\cite{extremewaves}. Each antenna is dual-polarized, so we have 64 antennas. The antenna RF input and outputs are connected through SMA connectors to the SDRs and operate at 28GHz, which is up/down-converted from/to the 4GHz baseband signal with a 24GHz UDC. We use an Artix-7-based FPGA~\cite{digilentCMOD} for programming the phased array registers and activating beam training commands using the SPI protocol. The registers are modified to set the phase and gain values (supports 6-bit phase and 6-bit gain values) obtained from MATLAB corresponding to a particular beam. Look Up Tables (LUTs) present in the phased arrays allow for storing the phase and gain values in a continuous register location that can be loaded rapidly without microcontroller intervention every time. The phased array is capable of performing a beam switch under 6$\mu$s. The FPGA, SDRs, and phased arrays are fed with a common 10MHz reference signal generated from the Ettus OctoClock clock distribution module that ensures sample level synchronization within the radio. A  power regulator board takes 12V input~\cite{battery} and distributes power to phased arrays, clock modules~\cite{adf5356} and fans. 

\textbf{$\blacksquare$ Base station setup:} For the base station, we use Ettus N310 USRP~\cite{usrpB210} that is connected to a PC through SFP optical cables. We operate the USRP at 4.05GHz center frequency and a sampling rate of 30.72MHz using GNU-Radio software~\cite{gnuradio}. We program the pre-calculated phase and gain values for the required angles for an experiment into the LUTs present in the phased arrays. When the USRP starts transmitting, a trigger is sent to the FPGA, which then activates the beam training, and the phased array loads the register values from the LUTs sequentially. 
\begin{figure*}[t!]
     \centering
     \begin{subfigure}[b]{0.32\textwidth}
        \centering
        \includegraphics[width=\linewidth]{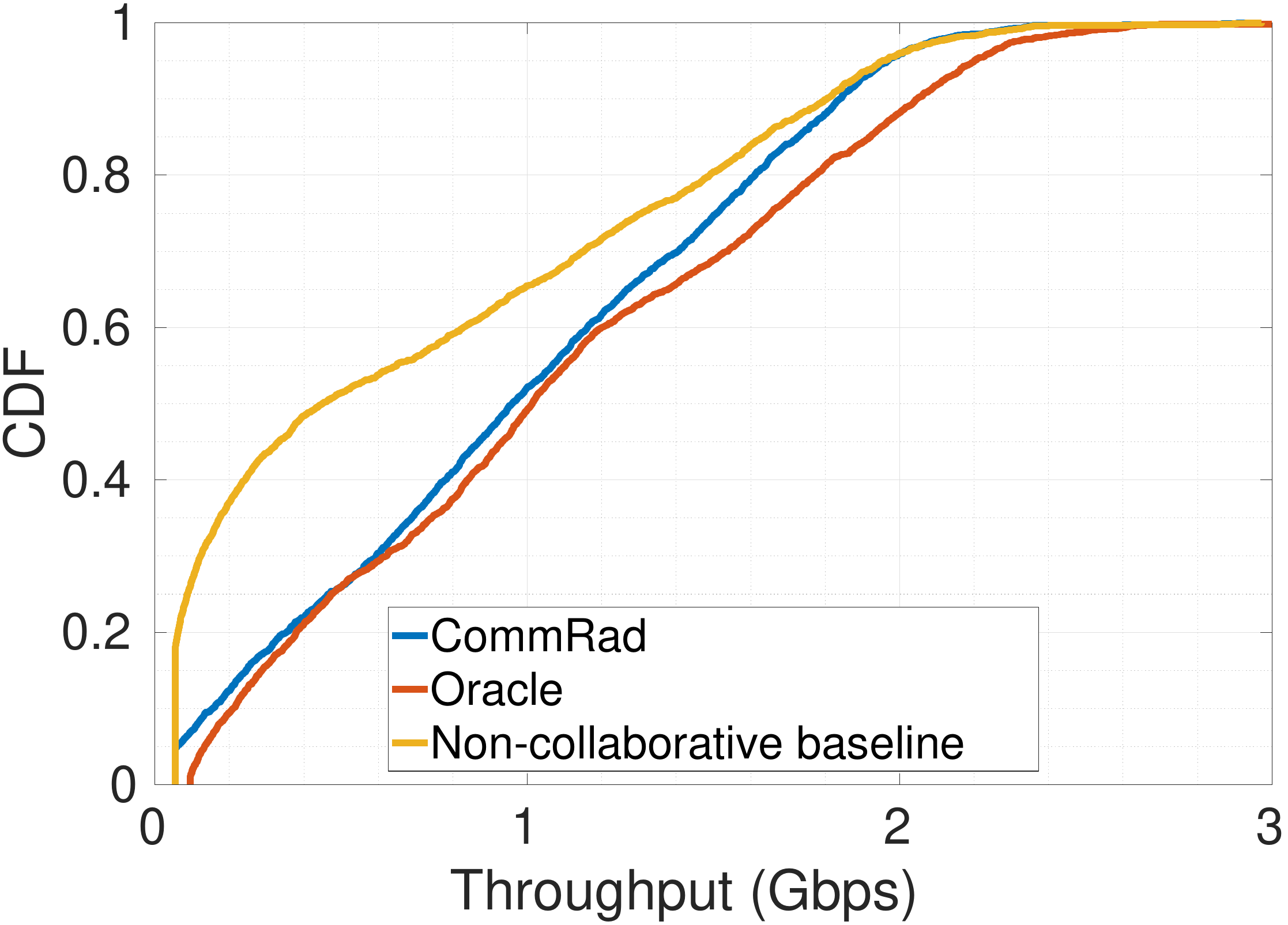}
        \caption{Throughput CDF}        
        \label{fig:thput}
     \end{subfigure}
      \hfill
     \begin{subfigure}[b]{0.32\textwidth}
        \centering
        \includegraphics[width=\linewidth]{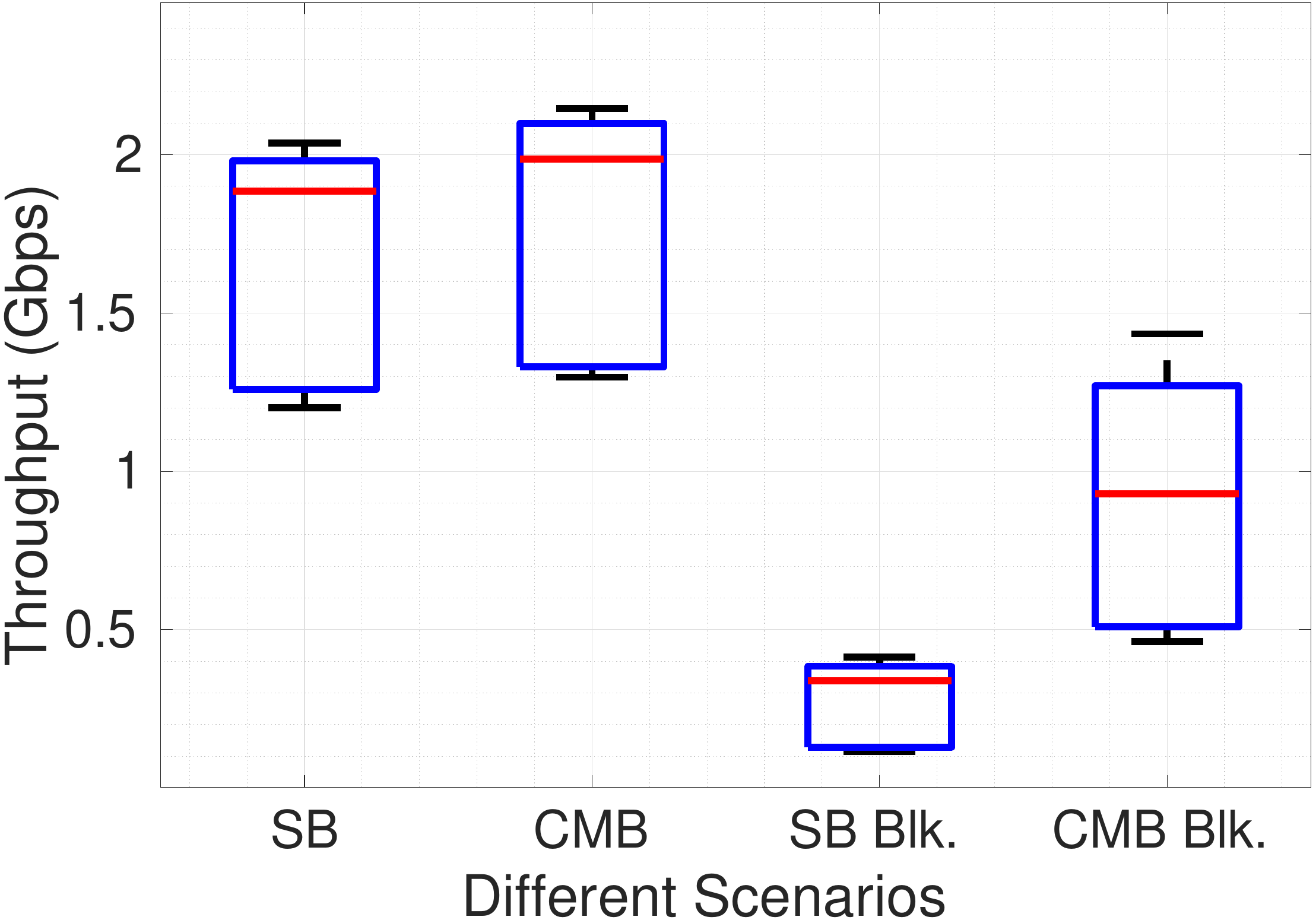}
        \caption{Throughput in different scenarios}
        \label{fig:cmb}        
     \end{subfigure}
  \hfill
       \begin{subfigure}[b]{0.28\textwidth}
        \centering
        \includegraphics[width=\linewidth]{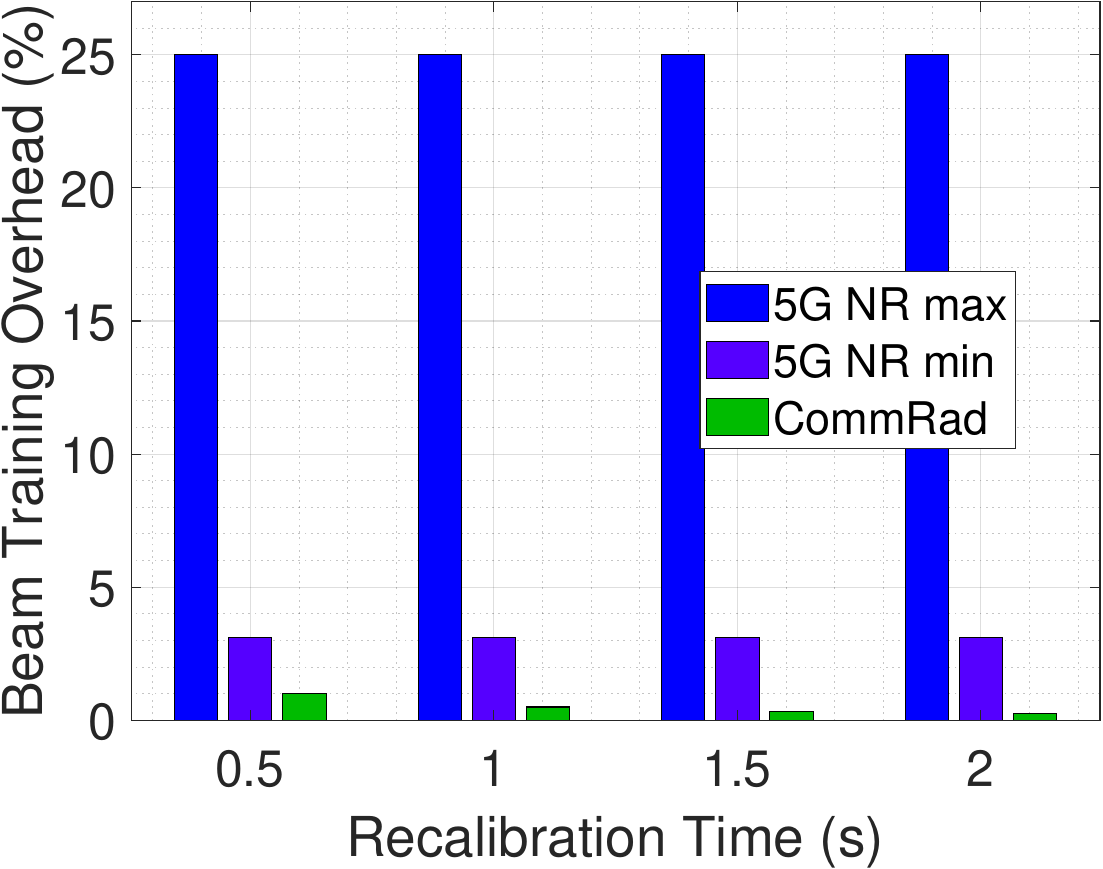}
        \caption{Overhead against recalibration time}        
        \label{fig:overhead}
     \end{subfigure}
        \caption{End-end results of \name: Improving median throughput by 2.5x compared to non-collaborative baseline. Supporting both single-beam (SB) and Constructive multi-beam (CMB) schemes with blockage mitigation. Reducing beam management overhead in 5GNR.}
        \label{fig:eval_figures}
\end{figure*}

\textbf{$\blacksquare$ User setup:} We developed a compact mobile setup for the user, consisting of a mmWave phased array, clock, Arduino micro-controller, a battery, and PlutoSDR. We choose plutoSDR~\cite{PlutoSDR} at the user because it is compact and allows streaming data directly from a USB stick without needing a laptop, thus making it easier to scale to multi-UE experiments. 
The phased arrays are controlled through an Arduino microcontroller to set a static quasi-omni-beam at each user. 

\textbf{$\blacksquare$ Waveform design:}
We use a modified 5G NR beam management protocol with a 12.5$\mu$s OFDM symbol with the preamble and data blocks. We choose this duration of OFDM symbol to get an integer number of samples at the sampling rate of 30.72MHz and allocate enough samples for the preamble. The preamble contains a typical 64-sample pseudorandom sequence, which is repeated twice and has a 32-sample cyclic prefix (CP). The preamble is followed by data that contains the repeated beam IDs that are also coded with a pseudorandom sequence to avoid the high Peak-to-Average-Power-Ratio (PAPR) prevalent in OFDM systems. 

\subsection{Radar Hardware and Software}
We used an INRAS RadarBook FMCW radar, which operates at 24GHz~\cite{radarbook2}. The radar is connected to the same PC as the base station with a 1 Gbit Ethernet cable. The radar has 2 TX and 8 RX antennas that are combined in post-processing to give a 16-antenna virtual array. The radar uses a bandwidth of 200MHz, with one chirp containing 1000 samples, with a slope of 2 MHz/$\mu$s having a ramp-up time of 100$\mu$s and a total duration of 200 $\mu$s. Each data frame contains 200 chirps, and one frame lasts 200 ms. The chirps are split equally between the two TX antennas, i.e., 100 chirps for each antenna. For the above parameters, we have a range resolution of 0.75m, velocity resolution of 0.75 m/s (14 Hz frequency resolution), and angular resolution of 7.16 degrees.

\textbf{$\blacksquare$ Synchronization and calibration:}
radio and radar each save timestamps using the NTP protocol during data collection for synchronization purposes. The radio achieves sample-level synchronization via the USRP, which sends triggers to the FPGA precisely when transmission or reception starts. This, in turn, triggers the phased array to begin beam-sweeping. By chaining triggers from the USRP to the FPGA to the phased array, we align the beam training with the initial sample transmission or reception. Figure \ref{fig:exp_setup} illustrates the clock and trigger signals used for synchronization and control. Additionally, calibration is necessary because the radar and radio, positioned about 15 cm apart, have different local coordinate systems. We establish a global coordinate system at the radio's location and apply a linear transform to convert radar points into this global coordinate system.

% Both radio and radar save timestamps according to the NTP protocol during the data collection. These timestamps are used to synchronize the data collected from radio and radar. Sample-level synchronization within the radio is obtained using USRP, which sends triggers to the FPGA at the exact time when transmission or reception starts, which in turn triggers the phased array to start beam-sweeping. By chaining triggers generated from the USRP to FPGA to phased array, we can align the start of the beam training with the first sample transmitter or received. Figure \ref{fig:exp_setup} describes the clock and trigger signals used for synchronization and control flow. Another important step is calibration: since the radar and radio are placed far apart (around 15 cm), the local coordinate system is different from each sensor. We define a global coordinate system at the radio's location and apply a linear transform to move radar points according to the global coordinate system.

%% file: 6_evaluation_v3.tex
% !TEX root = main.tex

\section{Evaluation}

We evaluate the tracking accuracy of \name and the resulting throughput and overhead reduction improvement in various indoor and outdoor scenarios as shown in Figure~\ref{fig:expt_scenario}. We performed multiple experiments and collected data for different scenarios containing varying numbers of users and different mobility patterns, with each experiment lasting for 10 to 15 seconds.

\textbf{$\blacksquare$ Oracle:} We consider Oracle as an entity that knows the best beam at all times for all users. The Oracle is obtained by performing a beam-sweep using the phased array at a very fast rate. The phased array scans 121 beams, with each beam lasting 12.5us. Thus, the total time required for a beam scan is 1.5ms. This procedure is executed for the entire duration of the experiment to capture the robust ground truth / Oracle data from the perspective of the radio. The Oracle perfectly captures the reflected beam when the direct path is blocked.

\textbf{$\blacksquare$ Baseline:} We implement a non-collaborative baseline that does not utilize the periodic contextual information from the radio for aiding radar tracking. To ensure fairness, we provide a one-time context for radar measurement at the beginning of tracking and let the radar track the user without context for the remaining duration of the experiment.
Here, the radar follows a similar approach for tracking by creating bounding boxes around the users after the radio gives the initial locations. Then, the radar tracks the user by updating the bounding box, but it does not recalibrate with the radio periodically. It essentially does the tracking on its own. We also compare the results with a reactive baseline, where the radio holds the beam until the beam is updated in the next beam training and has no input from the radar. 

We first present end-end results with \name and then benchmark individual components of \name separately.
\begin{figure*}[t!]
     \centering
     \begin{subfigure}[b]{0.24\textwidth}
        \centering
        \includegraphics[width=\linewidth]{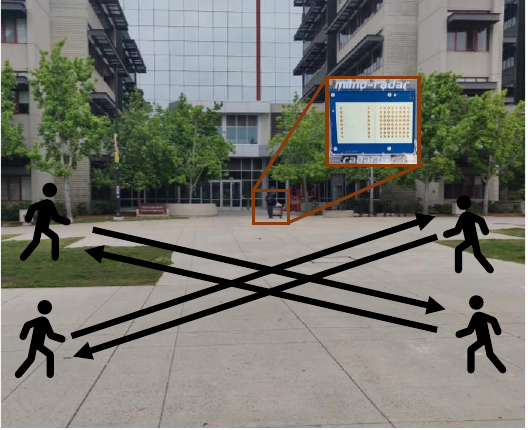}
        \caption{Crossing scenario}        
        \label{fig:scenario_crossing_4user}
     \end{subfigure}
     \hfill
        \begin{subfigure}[b]{0.20\textwidth}
        \centering
        \includegraphics[width=\linewidth]{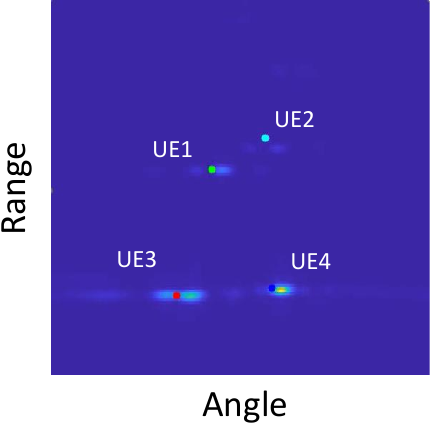}
        \caption{De-cluttered Radar View}
        \label{fig:radar_4user}        
     \end{subfigure}
     \hfill
     \begin{subfigure}[b]{0.25\textwidth}
        \centering
        \includegraphics[width=\linewidth]{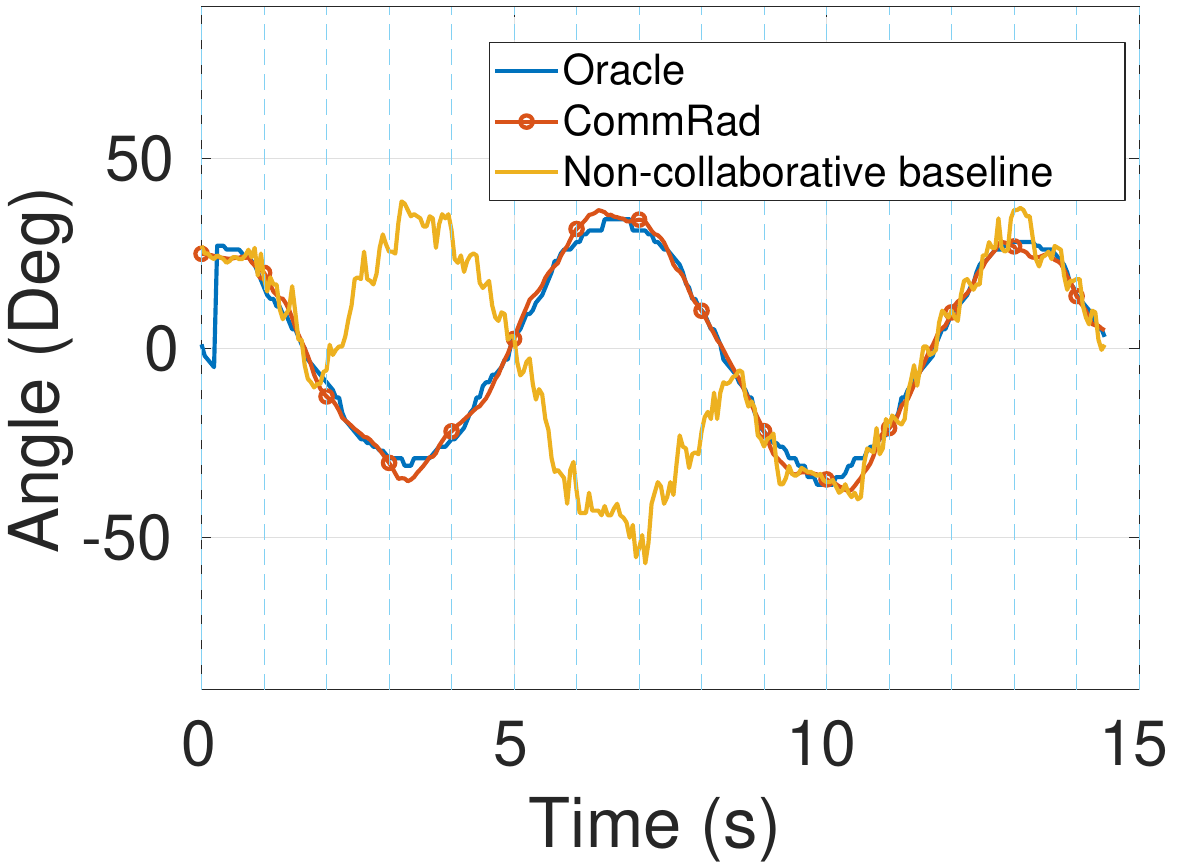}
        \caption{UE 1 Tracking}
        \label{fig:angle_time_user1_logid479}        
     \end{subfigure}
     \hfill
     \begin{subfigure}[b]{0.25\textwidth}
        \centering
        \includegraphics[width=\linewidth]{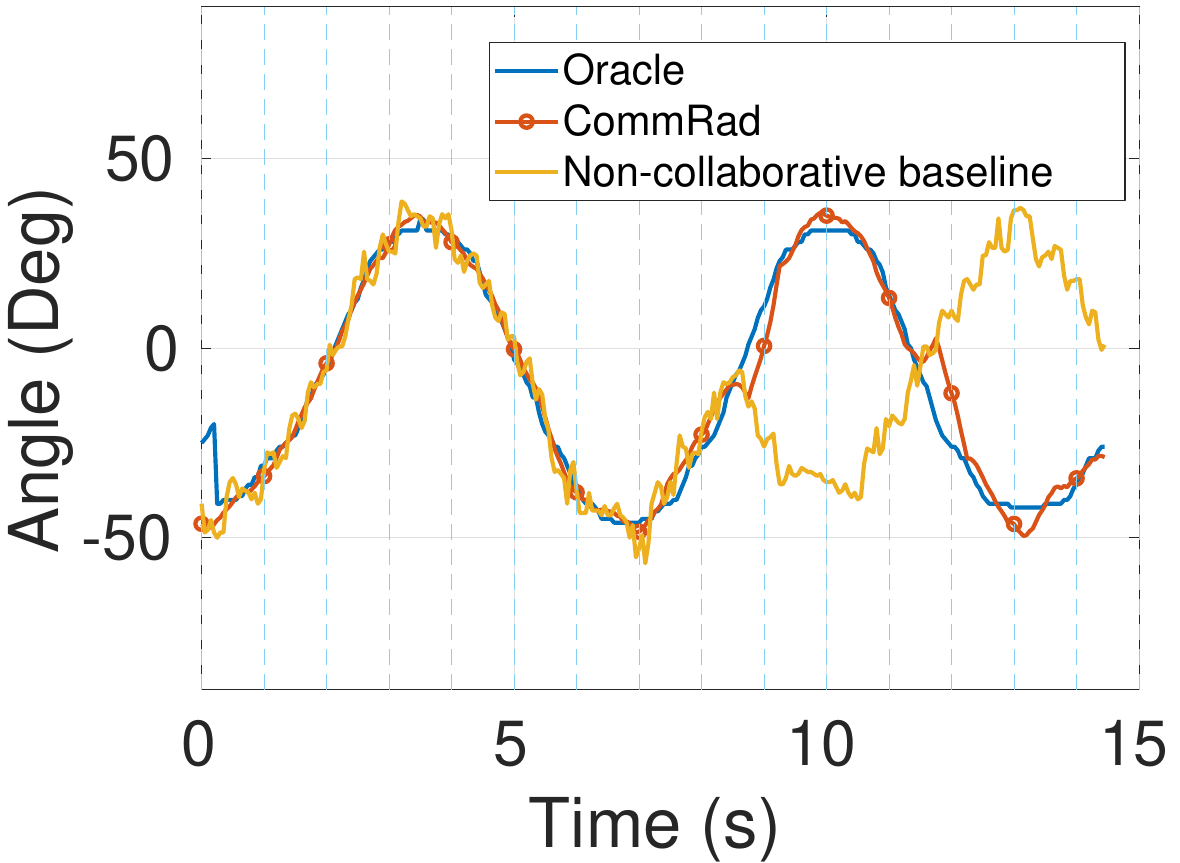}
        \caption{UE2 Tracking}
        \label{fig:angle_time_user2_logid479}        
     \end{subfigure}
        \caption{Scenario: Four users crossing each other. We remove the clutter and then show angle-tracking accuracy with collaborative learning. }
        \label{fig:scenario_crossing}
\end{figure*}

\subsection{Overall Throughput Improvement}
The goal of this evaluation is to show the throughput improvement of \name in diverse multi-user multipath environments. We obtain radio context information periodically, ensuring the radar can track and maintain beams in this duration.
We obtain a CDF of throughput measurements of \name and compare it against the non-collaborative baseline and Oracle in Figure~\ref{fig:eval_figures}(a). Here, we combined all the measurements across different environments and mobility patterns. The median throughput of \name is 1000 Mbps, while the baseline offers only 400 Mbps median throughput, thus providing a 2.5x throughput gain over the baseline while performing almost similar to the Oracle system.
The throughput improvement is more significant for the worst case, the 20th percentile of the cases, improving from 50 Mbps to 400 Mbps, providing a staggering 8x throughput gain. 
These results show that \name can help maintain a reliable link in difficult situations like device mobility and blockages. These improvements are due to robust radar tracking and blockage mitigation enabled by context awareness in \name's radar-radio collaborative framework.

Further, we evaluate the throughput improvement using traditional single-beam and constructive multi-beam techniques~\cite{jain2021two}. Constructive multi-beam uses two beams simultaneously, one in the direct path and the other in the reflected path direction, to provide reliable communication during blockages. The box plots in Figure \ref{fig:eval_figures}(b) show that the mean throughput increases by about 100 Mbps without blockers. With blockage, the traditional single-beam experiences a significant drop in throughput due to the direct LOS path being blocked. In contrast, the constructive multi-beam, which uses the reflected path to support the communication even during blockage to direct LOS path, experiences a drop in throughput by half, which is less significant compared to almost a 10x drop in the single beam case.

\begin{figure}[t!]
     \centering
     \begin{subfigure}[b]{0.33\textwidth}
        \centering
        \includegraphics[width=\linewidth]{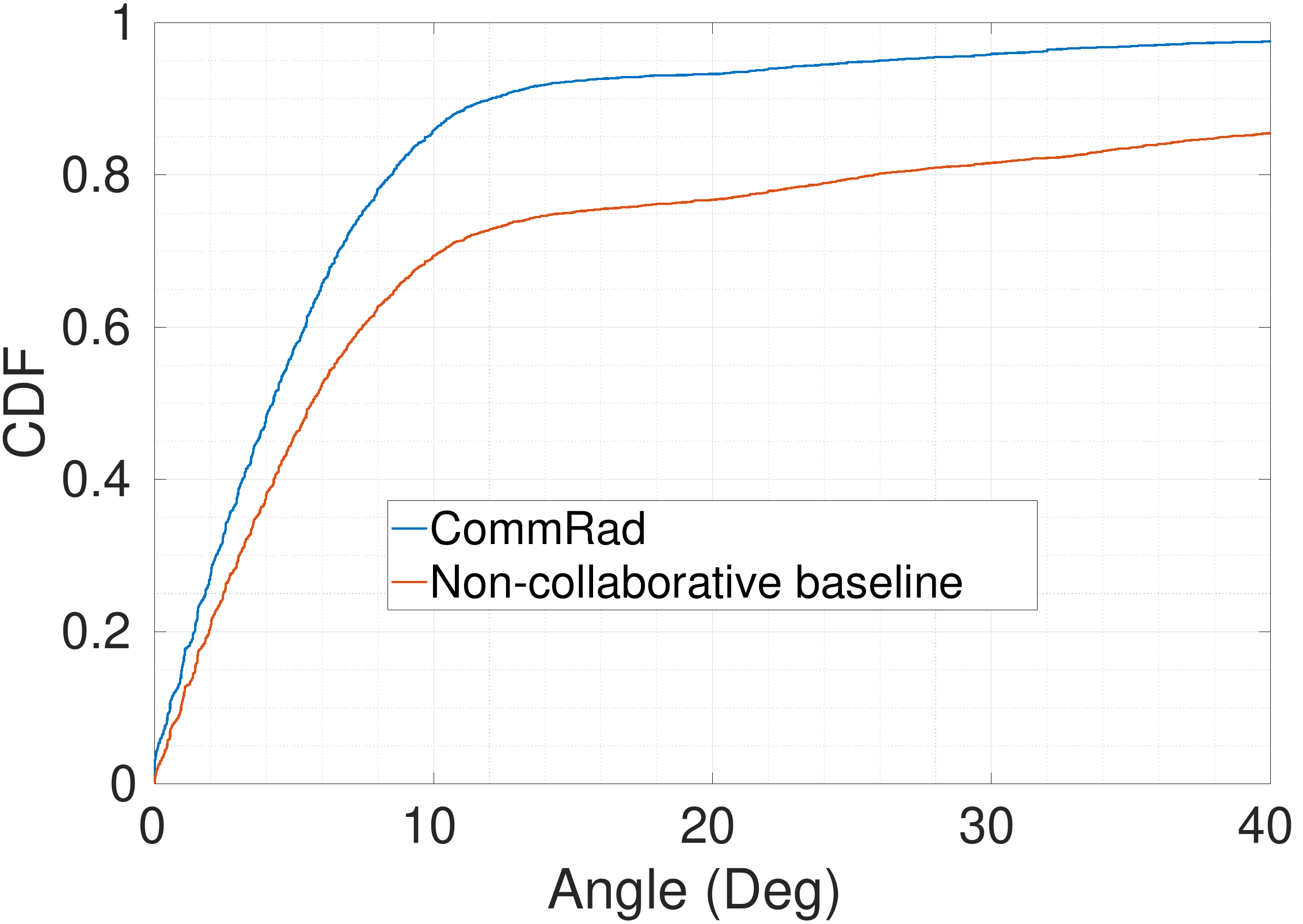}
        \caption{CDF for angle error}
        \label{fig:tracking_with_gt}        
     \end{subfigure}
    \hfill
     \begin{subfigure}[b]{0.33\textwidth}
        \centering
        \includegraphics[width=\linewidth]{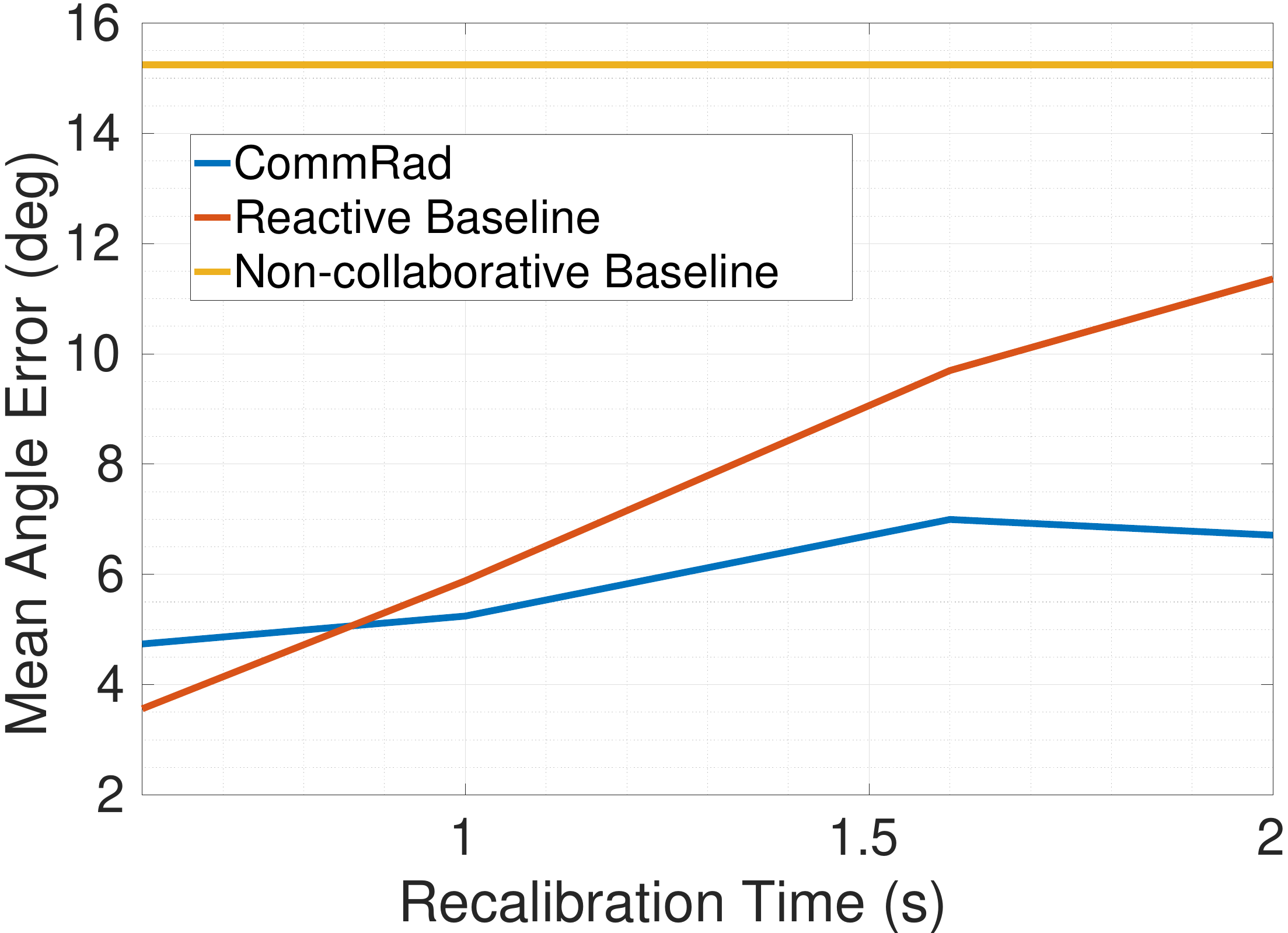}
        \caption{Angle error over recalibration time}
        \label{fig:angle_error_vs_recal_time}        
     \end{subfigure}
        \caption{\name's end-end angle tracking accuracy. }
        \label{fig:eval_figures_angle}
\end{figure}

\subsection{Overhead Reduction}
Next, we emphasize the advantage of using radar sensing to aid mmWave communication to reduce beam scan overhead. In cellular protocols like 5GNR, the base station (gNB) needs periodic beam scans and constant feedback for establishing and maintaining the beams towards the user, which creates a significant amount of overhead in terms of radio resources, affecting throughput and latency. Based on the periodicity of beam training in 5G NR, this overhead could be between 2.5\% to 25\%. \name can avoid this high overhead in many scenarios encountered in 5GNR. During initial access, the radar can indicate the arrival of a new object in the environment and its angle, thereby allowing gNB to scan only a small subset of angles. This can serve as a substitute for the beam refinement procedure suggested in 5G NR. During mobility, CommRad provides the optimal angle to the gNB, thereby reducing user feedback overhead. Finally, during blockages, while CommRad can provide alternate reflected paths, during the absence of strong reflectors, radar can provide the user location immediately without needing to re-establish the link. 
We show that \name significantly reduces this overhead down to 0.5\% when it relies on radar-based beam management with a periodicity of 0.5 seconds. At a higher periodicity of 2 seconds, this overhead is further reduced by 0.25\%. This shows \name as a low-overhead mechanism for the radio to spend more time on communication than spending time for beam scanning.

% \subsection{Impact of periodicity of beam scan}
% A major goal of this paper is to reduce the periodicity of beam scan and fill the gap by relying on radar more often and on radio beam scan less often. However, radar alone is not efficient for maintaining optimal beams because of sensing limitations, environment clutter and lack of semantic information about users and reflectors. Therefore, periodic calibration with radio is helpful. This periodic calibration if done at faster intervals would be accurate on the expense of high overhead. On the other hand, periodic calibration at lower intervals has low overhead, but could be less accurate. To show this tradeoff, we plot overall throughput with different periodicity of beam scan. We also plot beam scan overhead with different periodicity. 

\subsection{Angle Tracking Accuracy}
Accurate user tracking is the most important aspect of the system on which many features like reflector estimation, tracking, and blockage prediction impinge on. One of the main reasons for throughput gain for \name is because of its capability to track user angle accurately. 
We evaluate this by creating a link between the user and the base station radio and performing experiments where the user moves in various directions and speeds. We use radar to track the mobile user along with periodic updates from the radio. 

An example is shown in Figure~\ref{fig:scenario_crossing} with 4 users crossing each other diagonally. Here, we compare \name with a non-collaborative baseline comprising only radar tracking without collaboration with the radio. We observe that in complex scenarios where two users overlap, the non-collaborative baseline loses track of the user's identity and reports the same path for both users. With the periodic collaboration with radio, this effect is mitigated in \name.

To show the accuracy in user angle tracking, we plot the CDF of the error in angle prediction for \name compared to the non-collaborative baseline in Figure~\ref{fig:eval_figures_angle}(a). The angle error is the absolute value of the difference in Oracle angle and \name/baseline angle. We show that the median angular error of \name is 4 degrees while the baseline suffers from 8 degrees of median error. The 90th percentile error is more significant, where the \name error of 10 degrees is significantly better than the baseline error of more than 40 degrees.
Figure\ref{fig:eval_figures_angle}(b) shows the trade-off between angle error and recalibration time. The higher the recalibration time, the higher the angular error, consistent with the results expected from radar tracking objects without context for a long time. We also observe that around 2s, the angle error is around $7^o$, within the beamwidth of $10^o$ of the radio with 8 antennas. Therefore, an ideal trade-off between overhead reduction and reliable communication can be achieved when we update the user location at 2s intervals, causing the overhead reduction up to 0.25\% as estimated previously.

\begin{figure}[t!]
     \centering
     \begin{subfigure}[b]{0.24\textwidth}
        \centering
        \includegraphics[width=\linewidth]{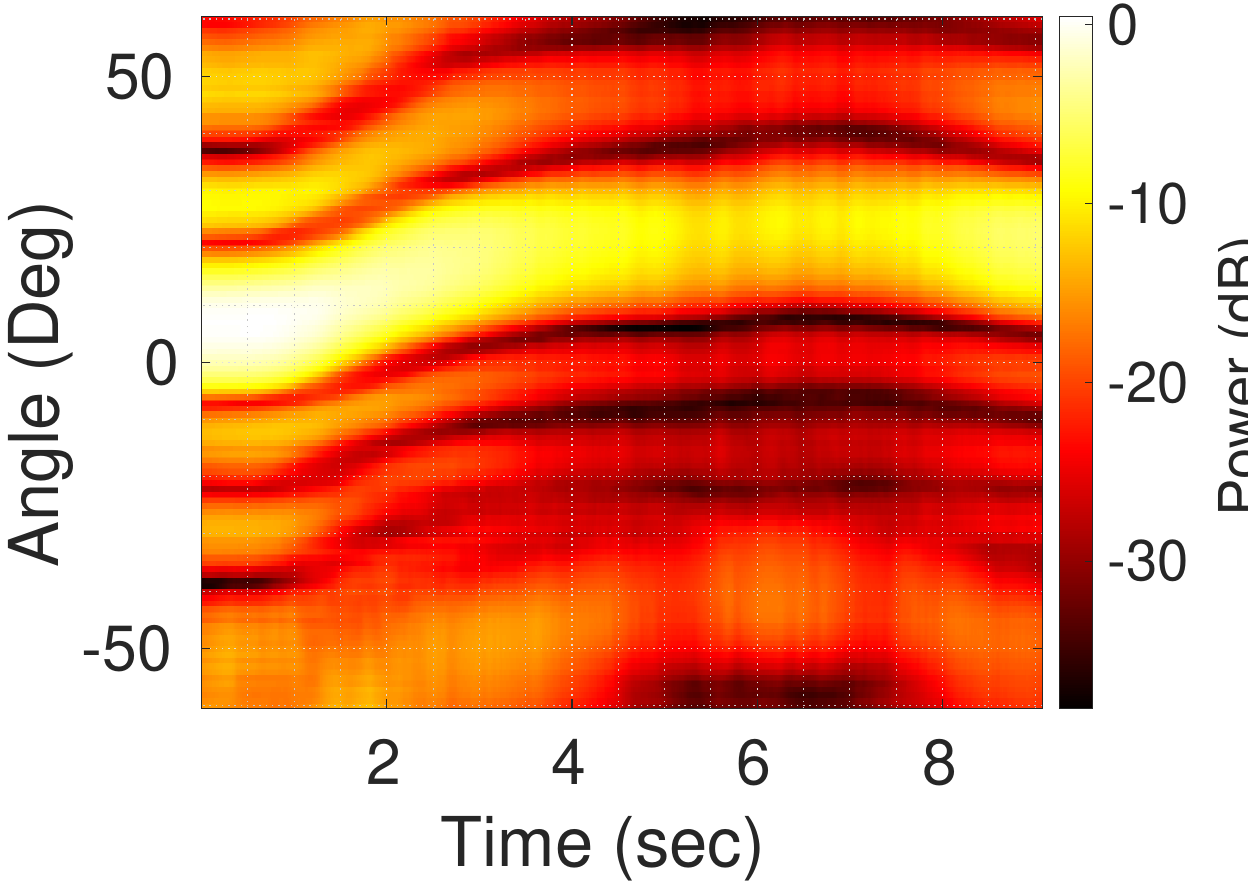}
        \caption{RSS power over time}        
        \label{fig:rss_heatmap_refl_logid407}
     \end{subfigure}
     \hfill
     \begin{subfigure}[b]{0.23\textwidth}
        \centering
        \includegraphics[width=\linewidth]{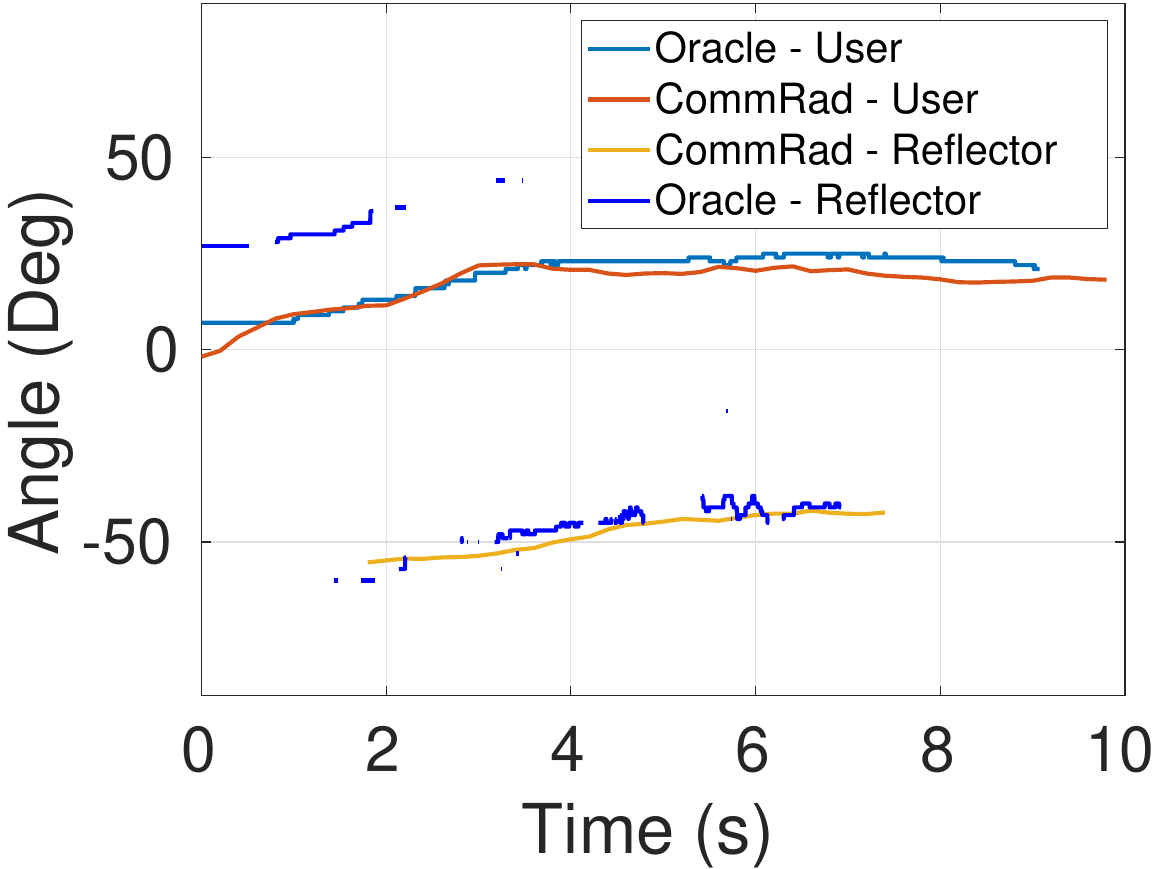}
        \caption{Angle with time}
        \label{fig:angle_time_refl_logid407}        
     \end{subfigure}
     \hfill
     % \begin{subfigure}[b]{0.28\textwidth}
     %    \centering
     %    \includegraphics[width=\linewidth]{figures/distance_time_refl_logid407.pdf}
     %    \caption{Distance with time}
     %    \label{fig:distance_time_refl_logid407}        
     % \end{subfigure}
        \caption{\name acquires ground truth by performing continuous beam training at all times. We show that the angle acquired by this beam scan (Oracle) matches the radar prediction.}
        \label{fig:rss_angle_tracking}
\end{figure}
\subsection{User/Reflector Tracking Accuracy}
% The radar  
We benchmark the accuracy of our radar tracking algorithm for both direct and reflected paths. For comparison, we obtain the ground truth angle through a continuous beam training process described earlier. Figure \ref{fig:rss_angle_tracking}(a) shows the RSS power across time and angle. This represents the ground truth (Oracle) information, where the high energy corresponds to the angle path traveled by the user over time. The second higher energy path corresponds to the reflected path seen by the radio. In Figure \ref{fig:rss_angle_tracking}(b), we compare \name's angle tracking with the Oracle angle and show that they match with high accuracy. The reflected path tracking is remarkable as the radar could accurately predict when the reflector is available to the user at the 2-sec mark and when it vanishes at the 7-sec mark due to user mobility. This high accuracy is achieved thanks to accurate context acquired by radar in reflector location, orientation, and size. 

% \name's estimates of LOS (User) and reflected path  overlaid on top of ground truth LOS and reflected path. We observe that CommRad predicts both LOS (User) and reflector paths accurately even when the user is moving and the angle is changing over time.
% Figure xx shows one snapshot of the range - angle of arrival profile obtained from the user with the high energy points classified as users, reflectors and blockers.

% \begin{figure*}[t!]
%     \centering
%     \includegraphics[width=\textwidth]{figures/recalibration.png}
%     \caption{How calibration with radio helps in accurately tracking two users using radar. In this scenario, radar is tracking two users shown in blue and red. The two user crossed their paths at 4 sec mark and after that the radio is confused user2 red path as user 1 path. Recalibration with radio has resolved this ambiguity. }
%     \label{fig:recalibration}
% \end{figure*}

% \subsection{Mobile reflector}
% One advantage of the proposed sensing-driven communication is the real-time tracking of environmental reflectors. When are the reflectors available for communication and when they vanish is useful for radio to develop strategies regarding the usage of those reflectors. If a reflector is no longer available then it won't be used for maintaining the communication link. We performed an experiment with different reflector mobility and derive the success rate of identifying whether the radar is successful in detecting reflector mobility and determine if it is still good for communication or not.

\subsection{Blockage Prediction and Mitigation}
\begin{figure}[t!]
     \centering
     \begin{subfigure}[b]{0.46\linewidth}
        \centering
        \includegraphics[width=\linewidth]{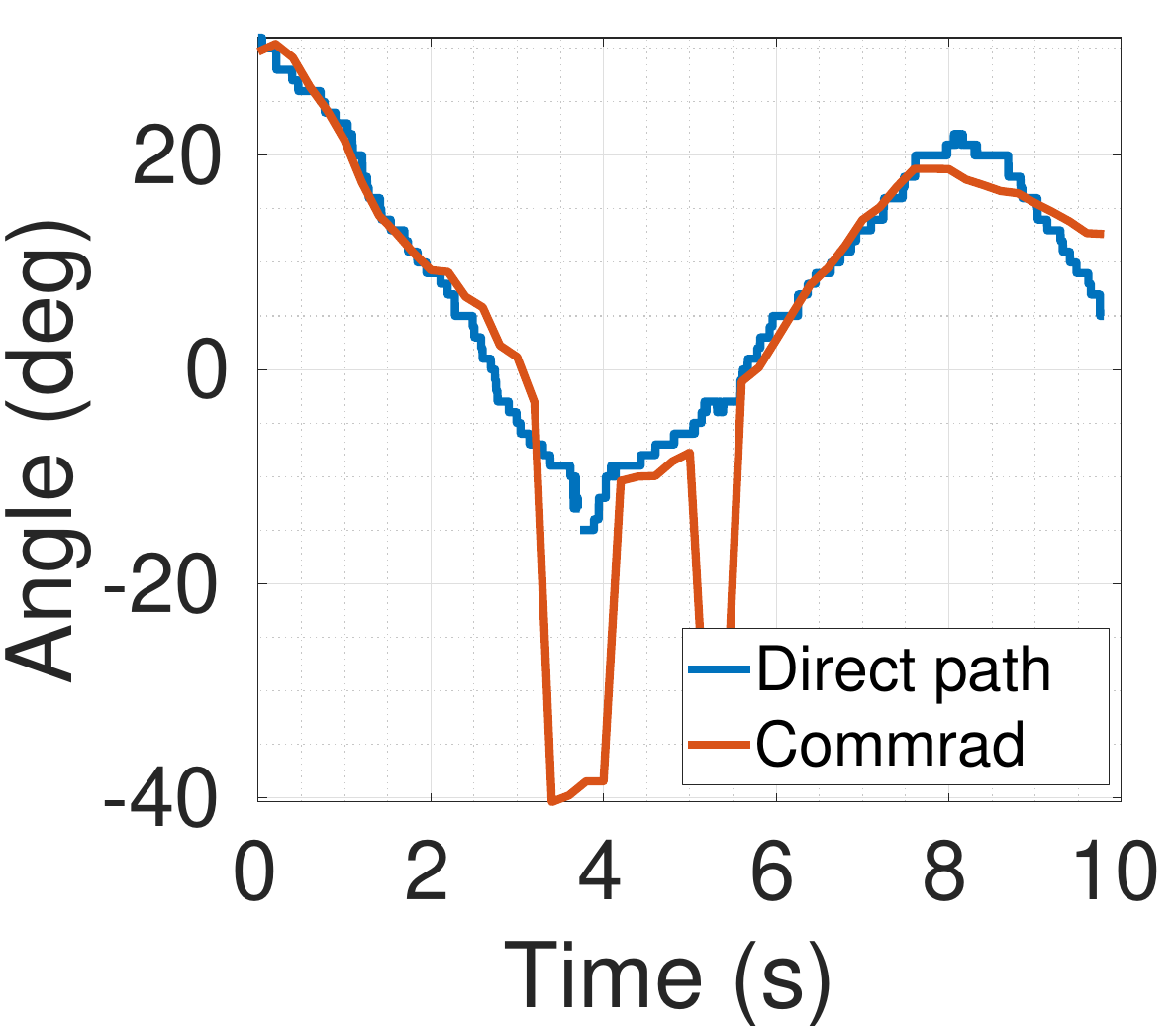}
        \caption{Angle adaptation}        
     \end{subfigure}
     \hfill
     \begin{subfigure}[b]{0.46\linewidth}
        \centering
        \includegraphics[width=\linewidth]{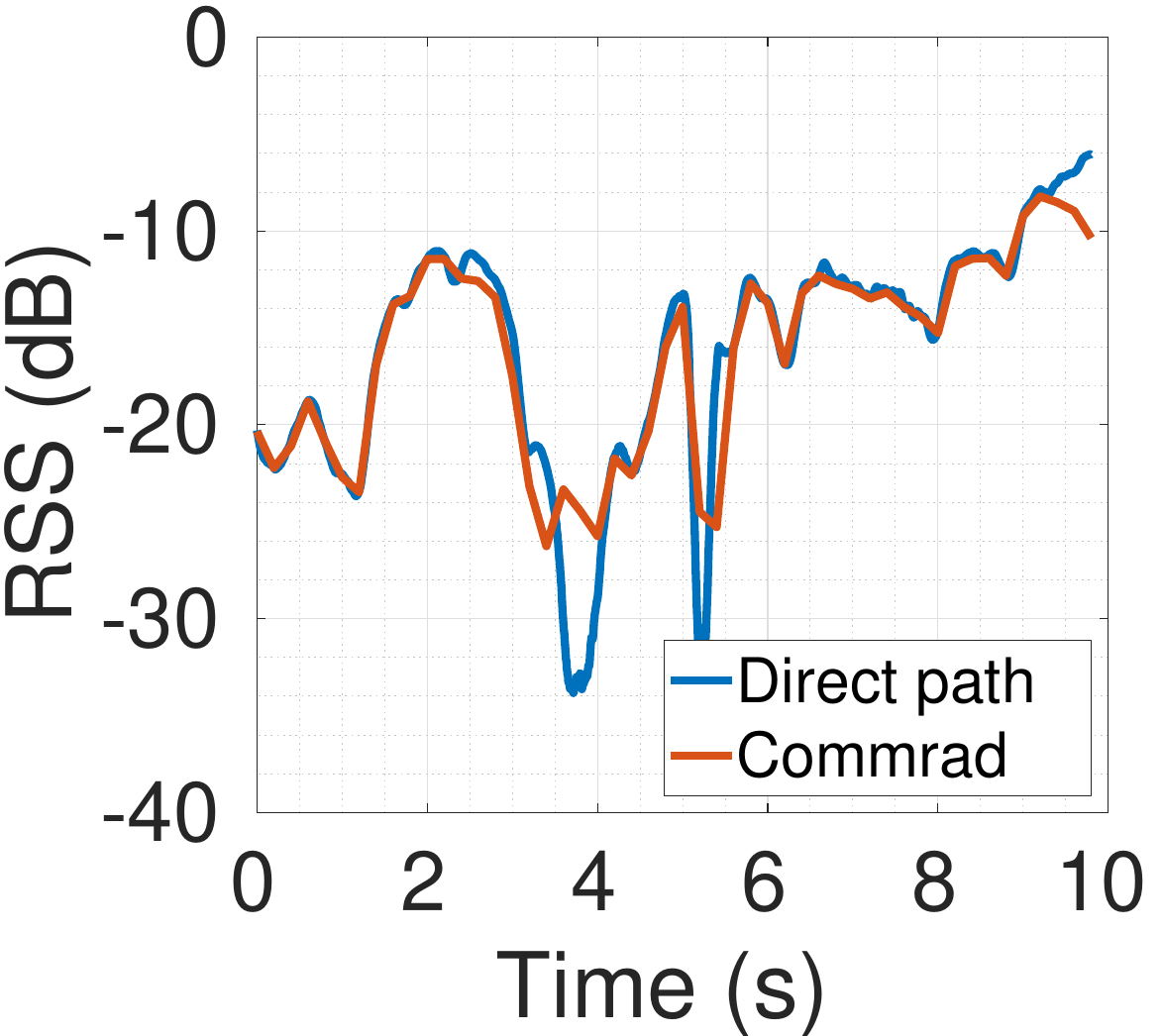}
        \caption{RSS Gain}
     \end{subfigure}
        \caption{\name adaptation to dynamic blockages.}
        \label{fig:blocker}
\end{figure}

Blockage for mmwave links reduces SNR and throughput and causes link outages. Utilizing alternate paths to maintain the link while the direct path is blocked is one of the proven solutions to prevent link outages. 
% However, due to uncertain blockage arrival and departure, it becomes tedious to proactively reach to blockage event. 
Since the potential reflected path can be anywhere in the entire angular space, it becomes tedious to do a full beam scan. Instead, we rely on radar to detect the blockage event and identify potential reflected paths to serve the user during the blockage. To demonstrate this, we show a 10-sec trace of adapted angles and corresponding RSS in Figures~\ref{fig:blocker}(a) and \ref{fig:blocker}(b), respectively. We compare \name's angle adaptation with a direct path baseline without any adaptation. As we can see, there is a sharp drop in angle twice at the 3-sec and 5-sec mark, which indicates a blockage event and beam switching to a reflected path. This adaptation in \name helped maintain high RSS, while the non-adaptive baseline suffered an RSS loss of more than 10 dB due to blockage. This shows the feasibility of radar-aided beam management and blockage adaptation in mmWave systems.

%% file: 8_discussion.tex
\section{Discussion and Future work}
We discuss potential limitations and future directions of \name:

% \textbf{Joint Sensing and Communication:} While \name deploys off-the-shelf radar and radio as separate hardware, they can be tightly integrated, sharing clocks and antennas to provide fine-grained synchronization and cost-effectiveness. Integrating hardware components is left as a future work. 
\textbf{$\blacksquare$ Sensor-fusion with contextual awareness:} \name deploys a 24 GHz Radar sensor in the ISM band to aid in beam management with contextual awareness. While Radars stand out as low-cost, privacy-preserving, and weather-resistant sensing modality, they can be fused with other sensors, such as radars at 60 GHz or 77 GHz ISM bands or light-based sensors such as cameras or lidars, to further improve beam management. Notably, \name's framework for contextual awareness can be generalized to other sensing modalities.

% \noindent
\textbf{$\blacksquare$ User-side beamforming:} We provide directional beam tracking at the base station side, assuming a Quasi-Omni directional user-side beam. This represents a typical scenario where the user side has a lower number of antennas due to power and size constraints for hand-held devices. Alternative techniques such as IMU-based user beam adaptation~\cite{wei2017pose} can be used along with \name. \\
\section{Conclusion}
In this paper, we investigate a collaborative bi-directional learning framework through the collaboration of radar and radio to improve next-generation mobile mmwave links. We identify complementary sensing information from mono-static radar and bi-static radio and combine them under our framework to get the best of both worlds. As a result, we improve mmwave link quality for mobile users even under environmental vulnerabilities such as blockages and beam misalignment due to mobility. 
% The future work would explore advanced algorithms such as machine learning together with collaborative learning to further improve the tracking accuracy, develop multi-modal sensing with additional sensors such as cameras, and explore more challenging scenarios such as extremely high vehicular mobility.

\noindent\textbf{Ethical concerns:} Does not raise any ethical issues.

\section{Acknowledgements}\label{sec:acks}
We are thankful to WCSNG lab members at UC San Diego for insightful feedback and to Raghav Subbaraman for their contribution to the initial phase of this project. This work was generously funded by Qualcomm Innovation Fellowship to Ish Kumar Jain and Raghav Subbaraman.

% We are grateful to the anonymous reviewers and our shepherd, Prof.
% % Romit Roy Choudhury, for their insightful feedback. We thank Prof.
% Gabriel Rebeiz, UCSD, and his group for the mmWave 5G
% phased arrays and transceivers. The research is supported by NSF CCRI \# 1925767.

%% file: main.bbl
\begin{thebibliography}{10}

\bibitem{3gpp}
{5G; Study on channel model for frequencies from 0.5 to 100 GHz}.
\newblock TR {138 901 V14.0.0}, {3rd Generation Partnership Project (3GPP)},
  2017-05.

\bibitem{fr2bands}
{5G NR Frequency Bands}.
\newblock \url{https://www.5gradio.com/5g-technology/5g-nr-frequency-bands/},
  September 2023.

\bibitem{abari2017enabling}
Omid Abari, Dinesh Bharadia, Austin Duffield, and Dina Katabi.
\newblock Enabling high-quality untethered virtual reality.
\newblock In {\em 14th $\{$USENIX$\}$ Symposium on Networked Systems Design and
  Implementation ($\{$NSDI$\}$ 17)}, pages 531--544, 2017.

\bibitem{5gnr}
3GPP.
\newblock {5G NR (New Radio) Release 16}. {3GPP}.
\newblock \url{https://www.3gpp.org/release-16}, Oct 2019.

\bibitem{zheng2023neuroradar}
Kai Zheng, Kun Qian, Timothy Woodford, and Xinyu Zhang.
\newblock Neuroradar: A neuromorphic radar sensor for low-power iot systems.
\newblock In {\em Proceedings of the 21st ACM Conference on Embedded Networked
  Sensor Systems}, pages 223--236, 2023.

\bibitem{cui2022integrated}
Kaiyan Cui, Qiang Yang, Leming Shen, Yuanqing Zheng, and Jinsong Han.
\newblock Integrated sensing and communication between daily devices and mmwave
  radars.
\newblock In {\em Proceedings of the 20th ACM Conference on Embedded Networked
  Sensor Systems}, pages 831--832, 2022.

\bibitem{bansal2020pointillism}
Kshitiz Bansal, Keshav Rungta, Siyuan Zhu, and Dinesh Bharadia.
\newblock Pointillism: accurate 3d bounding box estimation with multi-radars.
\newblock In {\em Proceedings of the 18th Conference on Embedded Networked
  Sensor Systems}, pages 340--353, 2020.

\bibitem{zheng2024enhancing}
Kai Zheng, Wuqiong Zhao, Timothy Woodford, Renjie Zhao, Xinyu Zhang, and Yingbo
  Hua.
\newblock Enhancing mmwave radar sensing using a phased-mimo architecture.
\newblock In {\em Proceedings of the 22nd Annual International Conference on
  Mobile Systems, Applications and Services}, pages 56--69, 2024.

\bibitem{dunna2023r}
Manideep Dunna, Kshitiz Bansal, Sanjeev~Anthia Ganesh, Eamon Patamasing, and
  Dinesh Bharadia.
\newblock R-fiducial: Millimeter wave radar fiducials for sensing traffic
  infrastructure.
\newblock In {\em 2023 IEEE 97th Vehicular Technology Conference
  (VTC2023-Spring)}, pages 1--7. IEEE, 2023.

\bibitem{demirhan2022radar}
Umut Demirhan and Ahmed Alkhateeb.
\newblock Radar aided 6g beam prediction: Deep learning algorithms and
  real-world demonstration.
\newblock In {\em 2022 IEEE Wireless Communications and Networking Conference
  (WCNC)}, pages 2655--2660. IEEE, 2022.

\bibitem{luo2023millimeter}
Hao Luo, Umut Demirhan, and Ahmed Alkhateeb.
\newblock Millimeter wave v2v beam tracking using radar: Algorithms and
  real-world demonstration.
\newblock {\em arXiv preprint arXiv:2308.01558}, 2023.

\bibitem{aydogdu2020distributed}
Canan Aydogdu, Fan Liu, Christos Masouros, Henk Wymeersch, and Mats
  Rydstr{\"o}m.
\newblock Distributed radar-aided vehicle-to-vehicle communication.
\newblock In {\em 2020 IEEE Radar Conference (RadarConf20)}, pages 1--6. IEEE,
  2020.

\bibitem{chen2023beam}
Leyan Chen, Kai Liu, Zhibo Zhang, and Baoqi Li.
\newblock Beam selection and power allocation: Using deep learning for
  sensing-assisted communication.
\newblock {\em IEEE Wireless Communications Letters}, 2023.

\bibitem{chen2023radar}
Zhanghua Chen and Jianxiao Xie.
\newblock Radar aided active prediction based on deep learning.
\newblock In {\em 2023 IEEE International Symposium on Broadband Multimedia
  Systems and Broadcasting (BMSB)}, pages 1--5. IEEE, 2023.

\bibitem{pearce2023multi}
Andre Pearce, J~Andrew Zhang, Richard Xu, and Kai Wu.
\newblock Multi-object tracking with mmwave radar: A review.
\newblock {\em Electronics}, 12(2):308, 2023.

\bibitem{ku2023characterizing}
Hansol Ku, Jinyue Song, Ding Zhang, Prasant Mohapatra, and Parth Pathak.
\newblock Characterizing real-time radar-assisted beamforming in mmwave v2v
  links.
\newblock In {\em 2023 20th Annual IEEE International Conference on Sensing,
  Communication, and Networking (SECON)}, pages 168--176. IEEE, 2023.

\bibitem{radarbook2}
Inras radarbook at 24 ghz and 77 ghz.
\newblock \url{https://inras.at/en/radarbook2/}, June 2024.

\bibitem{nabil2024beamwidth}
Yasser Nabil, Hesham ElSawy, Suhail Al-Dharrab, Hassan Mostafa, and Hussein
  Attia.
\newblock Beamwidth design tradeoffs in radar-aided millimeter-wave cellular
  networks: A stochastic geometry approach.
\newblock {\em IEEE Access}, 2024.

\bibitem{woodford2021spacebeam}
Timothy Woodford, Xinyu Zhang, Eugene Chai, Karthikeyan Sundaresan, and Amir
  Khojastepour.
\newblock Spacebeam: Lidar-driven one-shot mmwave beam management.
\newblock In {\em Proceedings of the 19th Annual International Conference on
  Mobile Systems, Applications, and Services}, MobiSys '21, page 389–401, New
  York, NY, USA, 2021. Association for Computing Machinery.

\bibitem{klautau2019lidar}
Aldebaro Klautau, Nuria Gonz{\'a}lez-Prelcic, and Robert~W Heath.
\newblock Lidar data for deep learning-based mmwave beam-selection.
\newblock {\em IEEE Wireless Communications Letters}, 8(3):909--912, 2019.

\bibitem{jiang2022lidar}
Shuaifeng Jiang, Gouranga Charan, and Ahmed Alkhateeb.
\newblock Lidar aided future beam prediction in real-world millimeter wave v2i
  communications.
\newblock {\em IEEE Wireless Communications Letters}, 12(2):212--216, 2022.

\bibitem{alrabeiah2020millimeter}
Muhammad Alrabeiah, Andrew Hredzak, and Ahmed Alkhateeb.
\newblock Millimeter wave base stations with cameras: Vision-aided beam and
  blockage prediction.
\newblock In {\em 2020 IEEE 91st Vehicular Technology Conference
  (VTC2020-Spring)}, pages 1--5. IEEE, 2020.

\bibitem{charan2023camera}
Gouranga Charan, Muhammad Alrabeiah, Tawfik Osman, and Ahmed Alkhateeb.
\newblock Camera based mmwave beam prediction: Towards multi-candidate
  real-world scenarios.
\newblock {\em arXiv preprint arXiv:2308.06868}, 2023.

\bibitem{imran2023environment}
Shoaib Imran, Gouranga Charan, and Ahmed Alkhateeb.
\newblock Environment semantic aided communication: A real world demonstration
  for beam prediction.
\newblock {\em arXiv preprint arXiv:2302.06736}, 2023.

\bibitem{ahmad2023vision}
Iftikhar Ahmad, Ahsan~Raza Khan, Rao Naveed~Bin Rais, Ahmed Zoha, Muhammad~Ali
  Imran, and Sajjad Hussain.
\newblock Vision-assisted beam prediction for real world 6g drone
  communication.
\newblock In {\em 2023 IEEE 34th Annual International Symposium on Personal,
  Indoor and Mobile Radio Communications (PIMRC)}, pages 1--7. IEEE, 2023.

\bibitem{charan2023user}
Gouranga Charan and Ahmed Alkhateeb.
\newblock User identification: A key enabler for multi-user vision-aided
  communications.
\newblock {\em IEEE Open Journal of the Communications Society}, 2023.

\bibitem{nie2023vision}
Jiali Nie, Quan Zhou, Junsheng Mu, and Xiaojun Jing.
\newblock Vision and radar multimodal aided beam prediction: Facilitating
  metaverse development.
\newblock In {\em Proceedings of the 2nd Workshop on Integrated Sensing and
  Communications for Metaverse}, pages 13--18, 2023.

\bibitem{wei2017pose}
Teng Wei and Xinyu Zhang.
\newblock Pose information assisted 60 {GHz} networks: Towards seamless
  coverage and mobility support.
\newblock In {\em Proceedings of the 23rd Annual International Conference on
  Mobile Computing and Networking}, pages 42--55. ACM, 2017.

\bibitem{va2015beam}
Vutha Va, Xinchen Zhang, and Robert~W Heath.
\newblock Beam switching for millimeter wave communication to support high
  speed trains.
\newblock In {\em 2015 IEEE 82nd Vehicular Technology Conference
  (VTC2015-Fall)}, pages 1--5. IEEE, 2015.

\bibitem{haider2018listeer}
Muhammad~Kumail Haider, Yasaman Ghasempour, Dimitrios Koutsonikolas, and
  Edward~W Knightly.
\newblock Listeer: Mmwave beam acquisition and steering by tracking indicator
  leds on wireless aps.
\newblock In {\em Proceedings of the 24th Annual International Conference on
  Mobile Computing and Networking}, pages 273--288. ACM, 2018.

\bibitem{sur2017wifi}
Sanjib Sur, Ioannis Pefkianakis, Xinyu Zhang, and Kyu-Han Kim.
\newblock {WiFi-assisted 60 {GHz} wireless networks}.
\newblock In {\em Proceedings of the 23rd Annual International Conference on
  Mobile Computing and Networking}, pages 28--41. ACM, 2017.

\bibitem{nitsche2015steering}
Thomas Nitsche, Adriana~B Flores, Edward~W Knightly, and Joerg Widmer.
\newblock Steering with eyes closed: mm-wave beam steering without in-band
  measurement.
\newblock In {\em 2015 IEEE Conference on Computer Communications (INFOCOM)},
  pages 2416--2424. IEEE, 2015.

\bibitem{jeong2015random}
Cheol Jeong, Jeongho Park, and Hyunkyu Yu.
\newblock Random access in millimeter-wave beamforming cellular networks:
  issues and approaches.
\newblock {\em IEEE Communications Magazine}, 53(1):180--185, 2015.

\bibitem{zhou2012efficient}
Liang Zhou and Yoji Ohashi.
\newblock Efficient codebook-based {MIMO} beamforming for millimeter-wave
  {WLANs}.
\newblock In {\em Personal Indoor and Mobile Radio Communications (PIMRC), 2012
  IEEE 23rd International Symposium on}, pages 1885--1889. IEEE, 2012.

\bibitem{barati2016initial}
C~Nicolas Barati, S~Amir Hosseini, Marco Mezzavilla, Thanasis Korakis,
  Shivendra~S Panwar, Sundeep Rangan, and Michele Zorzi.
\newblock Initial access in millimeter wave cellular systems.
\newblock {\em IEEE Transactions on Wireless Communications},
  15(12):7926--7940, 2016.

\bibitem{sur201560}
Sanjib Sur, Vignesh Venkateswaran, Xinyu Zhang, and Parmesh Ramanathan.
\newblock 60 {GHz} indoor networking through flexible beams: A link-level
  profiling.
\newblock In {\em ACM SIGMETRICS Performance Evaluation Review}, volume~43,
  pages 71--84. ACM, 2015.

\bibitem{hassanieh2018fast}
Haitham Hassanieh, Omid Abari, Michael Rodriguez, Mohammed Abdelghany, Dina
  Katabi, and Piotr Indyk.
\newblock Fast millimeter wave beam alignment.
\newblock In {\em Proceedings of the 2018 Conference of the ACM Special
  Interest Group on Data Communication}, pages 432--445. ACM, 2018.

\bibitem{aykin2019smartlink}
Irmak Aykin, Berk Akgun, and Marwan Krunz.
\newblock Smartlink: Exploiting channel clustering effects for reliable
  millimeter wave communications.
\newblock In {\em IEEE INFOCOM 2019-IEEE Conference on Computer
  Communications}, pages 1117--1125. IEEE, 2019.

\bibitem{heng2023grid}
Yuqiang Heng and Jeffrey~G Andrews.
\newblock Grid-free mimo beam alignment through site-specific deep learning.
\newblock {\em IEEE Transactions on Wireless Communications}, 23(2):908--921,
  2023.

\bibitem{wang2020demystifying}
Song Wang, Jingqi Huang, and Xinyu Zhang.
\newblock Demystifying millimeter-wave {V2X}: Towards robust and efficient
  directional connectivity under high mobility.
\newblock In {\em Proceedings of the 26th Annual International Conference on
  Mobile Computing and Networking}, pages 1--14, 2020.

\bibitem{heng2024site}
Yuqiang Heng, Yu~Zhang, Ahmed Alkhateeb, and Jeffrey~G Andrews.
\newblock Site-specific beam alignment in 6g via deep learning.
\newblock {\em arXiv preprint arXiv:2403.16186}, 2024.

\bibitem{zhou2012mirror}
Xia Zhou, Zengbin Zhang, Yibo Zhu, Yubo Li, Saipriya Kumar, Amin Vahdat, Ben~Y
  Zhao, and Haitao Zheng.
\newblock Mirror mirror on the ceiling: Flexible wireless links for data
  centers.
\newblock {\em ACM SIGCOMM Computer Communication Review}, 42(4):443--454,
  2012.

\bibitem{genc2010robust}
Zulkuf Genc, Umar~H Rizvi, Ertan Onur, and Ignas Niemegeers.
\newblock Robust 60 {GHz} indoor connectivity: Is it possible with reflections?
\newblock In {\em 2010 IEEE 71st vehicular technology conference}, pages 1--5.
  IEEE, 2010.

\bibitem{zhao2018improving}
Man Zhao, Danpu Liu, and Feng Yu.
\newblock Improving the robustness of 60 {GHz} indoor connectivity by
  deployment of mirrors.
\newblock In {\em 2018 IEEE 29th Annual International Symposium on Personal,
  Indoor and Mobile Radio Communications (PIMRC)}, pages 188--193. IEEE, 2018.

\bibitem{khawaja2018coverage}
Wahab Khawaja, Ozgur Ozdemir, Yavuz Yapici, Ismail Guvenc, and Yuichi
  Kakishima.
\newblock Coverage enhancement for mm wave communications using passive
  reflectors.
\newblock In {\em 2018 11th Global Symposium on Millimeter Waves (GSMM)}, pages
  1--6. IEEE, 2018.

\bibitem{zhu2014demystifying}
Yibo Zhu, Zengbin Zhang, Zhinus Marzi, Chris Nelson, Upamanyu Madhow, Ben~Y
  Zhao, and Haitao Zheng.
\newblock Demystifying 60{GHz} outdoor picocells.
\newblock In {\em Proceedings of the 20th annual international conference on
  Mobile computing and networking}, pages 5--16. ACM, 2014.

\bibitem{wei2017facilitating}
Teng Wei, Anfu Zhou, and Xinyu Zhang.
\newblock Facilitating robust 60 {GHz} network deployment by sensing ambient
  reflectors.
\newblock In {\em 14th $\{$USENIX$\}$ Symposium on Networked Systems Design and
  Implementation ($\{$NSDI$\}$ 17)}, pages 213--226, 2017.

\bibitem{zhou2019autonomous}
Anfu Zhou, Shaoyuan Yang, Yi~Yang, Yuhang Fan, and Huadong Ma.
\newblock Autonomous environment mapping using commodity millimeter-wave
  network device.
\newblock In {\em IEEE INFOCOM 2019-IEEE Conference on Computer
  Communications}, pages 1126--1134. IEEE, 2019.

\bibitem{zhou2020robotic}
Anfu Zhou, Shaoqing Xu, Song Wang, Jingqi Huang, Shaoyuan Yang, Teng Wei, Xinyu
  Zhang, and Huadong Ma.
\newblock Robotic millimeter-wave wireless networks.
\newblock {\em IEEE/ACM Transactions on Networking}, 28(4):1534--1549, 2020.

\bibitem{zhou2017beam}
Anfu Zhou, Xinyu Zhang, and Huadong Ma.
\newblock Beam-forecast: Facilitating mobile 60 {GHz} networks via model-driven
  beam steering.
\newblock In {\em IEEE INFOCOM 2017-IEEE Conference on Computer
  Communications}, pages 1--9. IEEE, 2017.

\bibitem{demirhan2022radarblock}
Umut Demirhan and Ahmed Alkhateeb.
\newblock Radar aided proactive blockage prediction in real-world millimeter
  wave systems.
\newblock In {\em ICC 2022-IEEE International Conference on Communications},
  pages 4547--4552. IEEE, 2022.

\bibitem{sur2016beamspy}
Sanjib Sur, Xinyu Zhang, Parmesh Ramanathan, and Ranveer Chandra.
\newblock Beamspy: enabling robust 60 {GHz} links under blockage.
\newblock In {\em 13th $\{$USENIX$\}$ Symposium on Networked Systems Design and
  Implementation ($\{$NSDI$\}$ 16)}, pages 193--206, 2016.

\bibitem{ganji2022terra}
Santosh Ganji, Jaewon Kim, and PR~Kumar.
\newblock Terra: blockage resilience in outdoor mm-wave networks.
\newblock In {\em Proceedings of the SIGCOMM'22 Poster and Demo Sessions},
  pages 55--57. 2022.

\bibitem{zhang2018mmchoir}
Ding Zhang, Mihir Garude, and Parth~H Pathak.
\newblock mmchoir: Exploiting joint transmissions for reliable 60{GHz} mmwave
  wlans.
\newblock In {\em Proceedings of the Eighteenth ACM International Symposium on
  Mobile Ad Hoc Networking and Computing}, pages 251--260. ACM, 2018.

\bibitem{sur2018towards}
Sanjib Sur, Ioannis Pefkianakis, Xinyu Zhang, and Kyu-Han Kim.
\newblock Towards scalable and ubiquitous millimeter-wave wireless networks.
\newblock In {\em Proceedings of the 24th Annual International Conference on
  Mobile Computing and Networking}, pages 257--271. ACM, 2018.

\bibitem{rappaport2013millimeter}
Theodore~S Rappaport, Shu Sun, Rimma Mayzus, Hang Zhao, Yaniv Azar, Kevin Wang,
  George~N Wong, Jocelyn~K Schulz, Mathew Samimi, and Felix Gutierrez.
\newblock Millimeter wave mobile communications for 5g cellular: It will work!
\newblock {\em IEEE access}, 1:335--349, 2013.

\bibitem{jain2021two}
Ish~Kumar Jain, Raghav Subbaraman, and Dinesh Bharadia.
\newblock Two beams are better than one: Towards reliable and high throughput
  mmwave links.
\newblock In {\em Proceedings of the 2021 ACM SIGCOMM 2021 Conference}, pages
  488--502, 2021.

\bibitem{shaw2013radiometry}
Joseph~A Shaw.
\newblock Radiometry and the friis transmission equation.
\newblock {\em American journal of physics}, 81(1):33--37, 2013.

\bibitem{kotaru2015spotfi}
Manikanta Kotaru, Kiran Joshi, Dinesh Bharadia, and Sachin Katti.
\newblock Spotfi: Decimeter level localization using wifi.
\newblock In {\em Proceedings of the 2015 ACM Conference on Special Interest
  Group on Data Communication}, pages 269--282, 2015.

\bibitem{extremewaves}
Extreme waves 5g phased arrays.
\newblock \url{https://www.extreme-waves.com/5g}, Jul 2023.

\bibitem{digilentCMOD}
Digilent cmod artix a7-35t fpga.
\newblock
  \url{https://digilent.com/reference/programmable-logic/cmod-a7/start}, Jul
  2023.

\bibitem{battery}
12v battery.
\newblock
  \url{https://www.amazon.com/65Watts-Portable-Charger-Battery-Notebooks/dp/B07P87JTVT},
  Jul 2023.

\bibitem{adf5356}
Analog devices inc., adf5356 module.
\newblock \url{https://www.analog.com/en/products/adf5356.html}, Jul 2023.

\bibitem{usrpB210}
Ettus usrp b210.
\newblock \url{https://www.ettus.com/all-products/ub210-kit/}, Jul 2023.

\bibitem{gnuradio}
Gnu radio.
\newblock \url{https://www.gnuradio.org/}, Jul 2023.

\bibitem{PlutoSDR}
Analog devices inc., adalm pluto sdr.
\newblock
  \url{https://www.analog.com/en/design-center/evaluation-hardware-and-software/evaluation-boards-kits/adalm-pluto.html},
  Jul 2023.

\end{thebibliography}
